\documentclass[11pt]{article}

\usepackage{graphicx}
\usepackage{amssymb}
\usepackage{color}
\usepackage{amsfonts}
\usepackage{amsmath}
\usepackage{natbib}
\usepackage{geometry}
\usepackage[bookmarks=false,colorlinks,citecolor=blue,urlcolor=blue]{hyperref}
\usepackage{setspace}

\newtheorem{theorem}{Theorem}[section]

\newtheorem{condition}{Condition}

\newtheorem{lemma}{Lemma}[section]

\newcommand{\ba}{\mbox{\bf a}}
\newcommand{\bb}{\mbox{\bf b}}
\newcommand{\bd}{\mbox{\bf d}}

\newcommand{\bu}{\mbox{\bf u}}
\newcommand{\bv}{\mbox{\bf v}}
\newcommand{\bw}{\mathbf{w}}

\newcommand{\bz}{\mbox{\bf z}}
\newcommand{\bA}{\mbox{\bf A}}
\newcommand{\bB}{\mbox{\bf B}}

\newcommand{\bI}{\mbox{\bf I}}

\newcommand{\bM}{\mbox{\bf M}}
\newcommand{\bQ}{\mbox{\bf Q}}
\newcommand{\bR}{\mbox{\bf R}}
\newcommand{\bS}{\mbox{\bf S}}

\newcommand{\bX}{\mbox{\bf X}}

\newcommand{\bY}{\mbox{\bf Y}}

\newcommand{\bzero}{\mbox{\bf 0}}
\newcommand{\bvarepsilon}{\mbox{\boldmath $\varepsilon$}}

\newcommand{\bbeta}{\mbox{\boldmath $\beta$}}

\newcommand{\bpsi}{\mbox{\boldmath $\psi$}}

\newcommand{\bxi}{\mbox{\boldmath $\xi$}}

\newcommand{\argmin}{\mbox{argmin}}

\newcommand{\MN}{\mbox{MN}}

\onehalfspace

\begin{document}

\title{P{\Large{ENALIZED}} C{\Large OMPOSITE} Q{\Large UASI}-L{\Large IKELIHOOD} \\ {\Large FOR} U{\Large LTRAHIGH}-D{\Large IMENSIONAL} V{\Large ARIABLE} S{\Large ELECTION}  }
\author{B{\footnotesize{Y}} J{\footnotesize{ELENA}} B{\footnotesize{RADIC}}$^\dag$\thanks{
This research was partially supported by NSF Grants DMS-0704337 and DMS- 0714554 and NIH Grant R01-GM072611.  The bulk of the work was conducted while Weiwei Wang was a postdoctoral fellow at Princeton University.},  \  J{\footnotesize{IANQING}} F{\footnotesize{AN}}$^\dag$ {\footnotesize{AND}} W{\footnotesize{EIWEI}} W{\footnotesize{ANG}}$^\ddag$ \\ {\em $\dag$ Department of Operations Research and Financial Engineering} \\ {\em Princeton University} \\ {\em and } \\ {\em$\ddag$ Biostatistics/Epidemiology/Research Design (BERD) Core}\\ {\em University of Texas Health Science} {\em Center at Houston}}

\date{J{\footnotesize{UNE}} 30, 2010}
\maketitle


\begin{abstract}
In  high-dimensional model selection problems, penalized
least-square approaches have been extensively  used.  This
paper addresses the question of  both robustness and efficiency of
penalized model selection methods, and proposes a data-driven weighted linear combination of convex
loss functions, together with weighted $L_1$-penalty. It is completely data-adaptive and does not require prior knowledge of the error distribution. The weighted $L_1$-penalty is used both to ensure the
convexity of the penalty term and to ameliorate the bias caused by the $L_1$-penalty. In the setting with dimensionality much larger than the sample size,  we establish a strong oracle property of the
proposed method that possesses both the model selection consistency and estimation efficiency for the true non-zero coefficients. As specific examples, we introduce  a robust method of composite L1-L2, and
optimal composite quantile method and evaluate their performance in both simulated and real data examples.

\vskip 10pt




\end{abstract}

\newpage

\section{Introduction}\label{sec1}

Feature extraction and model selection are important for sparse high
dimensional data analysis in many research areas  such as genomics,
genetics and machine learning.  Motivated by the need of robust and
efficient high dimensional model selection method, we introduce a new penalized quasi-likelihood estimation for linear model with high dimensionality of parameter space.

Consider the estimation of the unknown parameter  $\bbeta$  in the  linear regression model
\begin{equation}\label{eq1}
\bY = \bX \bbeta + \bvarepsilon,
\end{equation}
where $\bY = (Y_1,\cdots,Y_n)^T$ is an $n$-vector of response, $\bX = (\bX_1, \cdots, \bX_n)^T$ is an $n \times p$ matrix of independent variables with  $\bX_i^T$ being its $i$-th row, $\bbeta = (\beta_1,...,\beta_p)^T$ is a $p$-vector of
unknown parameters and $\bvarepsilon = (\varepsilon_1,...,\varepsilon_n)^T$ is an $n$-vector of i.i.d. random errors with mean zero, independent of $\bX$. When the dimension $p$ is high it is commonly assumed that only a
small number of predictors actually contribute to the response vector
$\bY$, which leads to the sparsity pattern in the unknown parameters
and thus makes variable selection crucial.  In many applications such as
genetic association studies and disease classifications using high-throughput data such as microarrays with gene-gene interactions, the number of variables $p$ can be much larger than the sample size
$n$.  We will refer to such  problem as ultrahigh-dimensional problem and model it by assuming $\log p = O(n^\delta)$ for some $\delta \in (0, 1)$.  Following \citet{FL09}, we will refer to $p$
as a non-polynomial order or NP-dimensionality for short.

Popular approaches such as LASSO \citep{T96}, SCAD \citep{FL01}, adaptive LASSO \citep{Z06} and elastic-net \citep{ZZ09} use penalized least-square
regression:
\begin{equation} \label{eq2}
\hat{\bbeta} = \arg \min_{\boldsymbol\beta} \sum_{i=1}^n \left( Y_i - \bX_i^T \bbeta \right )^2 +n \sum_{j=1}^p p_\lambda(|\beta_j|).
\end{equation}
where $p_{\lambda}(\cdot)$ is a specific penalty function. The quadratic loss is popular  for its mathematical beauty but is not robust to non-normal errors and presence of outliers. Robust regressions such as the least absolute deviation and quantile regressions have recently been used in variable selection techniques when $p$ is finite \citep{WU09, ZY08, LZ08}. Other possible choices
of robust loss functions  include Huber's loss \citep{H64}, Tukey's bisquare, Hampel's psi, among others. Each of these loss functions performs well under a certain class of error
distributions:  quadratic loss is suitable  for normal distributions,
least absolute deviation  is suitable for heavy-tail distributions and is the most efficient for double exponential distributions, Huber's loss performs
well for contaminated normal distributions. However, none of them is universally better than all others.  How to construct an adaptive loss function that is applicable to  a large collection of error distributions?

We propose a simple and yet effective  quasi-likelihood function, which replaces the quadratic loss by a weighted linear combination of convex loss functions:
\begin{equation} \label{eq3}
     \rho_{\mathbf{w}} = \sum_{k=1}^{K} w_k \rho_k,
\end{equation}
where $\rho_1,...,\rho_K$ are convex loss functions and $w_1,...,w_K$ are positive constants chosen to minimize the asymptotic variance of the resulting estimator. From the point of view of nonparametric statistics, the functions $\{\rho_1, \cdots,
\rho_K \}$ can be viewed as a set of basis functions, not necessarily orthogonal, used to approximate the unknown log-likelihood function of the error distribution.  When the set of loss functions is
large, the quasi-likelihood function can well approximate the log-likelihood function and therefore yield a nearly efficient method. This kind of ideas appeared already in  traditional statistical
inference  with finite dimensionality \citep{K84,BRW92}. We will extend it to the sparse statistical inference with NP-dimensionality.

The quasi-likelihood function  \eqref{eq3} can be directly used
together with any penalty function such as $L_p$-penalty with $0<p <1$
\citep{FF93},  LASSO i.e. $L_1$-penalty \citep{T96}, SCAD
\citep{FL01}, hierarchical penalty \citep{BRT08}, resulting in the penalized composite quasi-likelihood problem:
\begin{equation} \label{eq4}
 \min_{\boldsymbol\beta} \sum_{i=1}^n \rho_{\bw} ( Y_i - \bX_i^T \bbeta ) +n \sum_{j=1}^p p_\lambda(|\beta_j|).
\end{equation}
Instead of using folded-concave penalty functions, we use the weighted $L_1$- penalty of the form
$$
n \sum_{j=1}^p \gamma_\lambda (|\beta_j^{(0)}|) |\beta_j|
$$
for some function $\gamma_\lambda$ and initial estimator $\bbeta^{(0)}$, to ameliorate the bias in $L_1$-penalization \citep{FL01, Z06, FL09} and to maintain the  convexity  of the problem.  This leads to the following
convex optimization problem:
\begin{equation}\label{eq5}
\widehat{\bbeta}_{\bw} = \arg\!\min_{\boldsymbol\beta} \sum_{i=1}^n \rho_{\mathbf{w}} \left(Y_i - \bX_i^T \bbeta \right) + n \sum_{j=1}^p \gamma_\lambda (|\beta^{(0)}_j|)|\beta_j|
\end{equation}
When $\gamma_\lambda(\cdot) = p_\lambda'(\cdot)$, the derivative of the penalty function, (\ref{eq5}) can be regarded as the local linear approximation to  problem (\ref{eq4})
\citep{ZL08}.  In particular, LASSO \citep{T96} corresponds to
$\gamma_\lambda(x)= \lambda$, SCAD reduces to
\citep{FL01}
\begin{equation}\label{eq6}
 \gamma_\lambda (x)= \lambda \{ I(x \leq \lambda) + \frac{(a \lambda-x)_+}{(a-1)\lambda}I(x> \lambda)\},
\end{equation}
and adaptive LASSO \citep{Z06} takes $\gamma_\lambda (x)= \lambda |x|^{-a}$ where $a>0$.

There is a rich literature in establishing the oracle property for penalized regression methods, mostly for large but fixed $p$
\citep{FL01,Z06,YL07,ZY08}. One of the early papers on diverging $p$ is the work by
\cite{FP04} under conditions of $p=O(n^{1/5})$. More recent works of the similar kind include \cite{HHM08}, \cite{ZZ09},
\cite{XH09}, which assume that the number of non-sparse elements $s$ is finite. When the dimensionality $p$ is of polynomial order, \cite{KCO08} recently  gave the conditions under which the SCAD estimator is an oracle estimator.  We would like to further  address this problem when $\log p = O(n^\delta)$ with
$\delta \in (0, 1)$ and  $s=O(n^{\alpha_0})$ for $\alpha_0 \in (0,1)$, that is when the dimensionality is of exponential order.

The paper is organized as follows. Section \ref{sec2} introduces  an
easy to implement two-step computation procedure. Section \ref{sec3} proves the strong oracle property of the  weighted $L_1$-penalized quasi-likelihood approach with discussion on the choice of weights and corrections for convexity.
Section \ref{sec4} defines  two specific instances of the proposed
approach and compares their asymptotic efficiencies. Section
\ref{sec5} provides a comprehensive simulation study as well as a real data example of the SNP selection for the Down syndrome. Section \ref{sec6} is devoted to the discussion. To facilitate the readability, all the proofs are relegated to the Appendices A, B \& C.

\section{Penalized adaptive composite quasi-likelihood}\label{sec2}

We would like to describe the proposed two-step adaptive computation procedure and defer the justification of the appropriate choice of the weight vector $\mathbf{w}$ to Section \ref{sec3}.

In the first step, one will get the initial estimate
$\hat{\bbeta}^{(0)}$ using  the
 LASSO procedure, i.e:
\[
 \hat{\bbeta}^{(0)} = \arg\!\min_{\boldsymbol{\beta}} \sum_{i=1}^n  \left(Y_i - \bX_i^T \bbeta \right)^2 +  n \lambda \sum_{j=1}^p |\beta_j|.
\]
and estimate the residual vector $\bvarepsilon^0 = \bY - \bX
\widehat{\bbeta}^{(0)}$ (for justification see discussion following Condition \ref{cond2}).  The matrix $\mathbf M$ and vector $\ba$
are calculated as follows:
\[
{\mathbf M}_{kl}= \frac{1}{n} \sum_{i=1}^n \psi_k({\varepsilon}_i^0)\psi_l({\varepsilon}_i^0), \ \mbox{and}  \ \ \ a_k= \frac{1}{n} \sum_{i=1}^n \partial \psi_k({\varepsilon}_i^0), \ \ \ (k,l=1,...,K),
\]
where $\psi_k(t)$ is a choice of the subgradient of $\rho_k(t)$,
${\varepsilon}_i^0$ is the $i$-th component of $\bvarepsilon^0$, and
$a_k$ should be considered as a consistent estimator of $E \partial \psi_k (\varepsilon)$, which is the derivative of
$E  \psi_k (\varepsilon+c)$ at $c = 0$.  For example, when
$\psi_k (x) = \mbox{sgn}(x)$, then $E \psi_k(\varepsilon + c) = 1
- 2 F_\varepsilon(-c)$ and $E \partial \psi_k (\varepsilon) = 2
f_\varepsilon (0)$.   The optimal weight is then determined as
\begin{equation}  \label{eq7}
\bw_{opt} =  \argmin_{\bw \geq 0, \ba^T \bw = 1} \mathbf{w}^T\mathbf{M}\mathbf{w}.
\end{equation}
In the second step,  one calculates the quasi maximum likelihood estimator (QMLE) using weights ${\mathbf{w}}_{opt}$ as
\begin{equation} \label{eq8}
\widehat{\bbeta}^{\footnotesize{\mbox{a}}} = \arg\min_{\boldsymbol{\beta}} \sum_{i=1}^n \rho_{\mathbf{w}_{opt}} \left( Y_i - \bX_i^T \bbeta \right) + n \sum_{j=1}^p \gamma_\lambda
(|\hat{\beta}^{(0)}_j |)|\beta_j|.
\end{equation}
{\bf Remark  {\color{red}1}}:
Note that zero is not an absorbing state in the minimization problem \eqref{eq8}.  Those elements that are estimated as zero in the initial estimate $\bbeta^{(0)}$ have a chance to escape from zero, whereas those nonvanishing elements can be estimated as zero in \eqref{eq8}.

\noindent{\bf Remark  {\color{red}2}}: The number of loss functions $K$ is typically small or moderate in practice.
 Problem \eqref{eq7} can be easily solved using a quadratic programming algorithm.  The resulting vector $\bw_{opt}$ can have vanishing
components,  automatically eliminating inefficient loss functions in the
second  step (\ref{eq8}) and hence learning the best approximation of the unknown log-likelihood function.
This can lead to considerable computational gains. See Section~\ref{sec4} for additional details.

\noindent{\bf Remark  {\color{red}3}}: Problem (\ref{eq8}) is a convex optimization problem when
$\rho_k$'s are all convex and  $\gamma_\lambda (|\hat{\beta}^{(0)}_j
|)$ are all nonnegative. This class of problems can be solved with fast and
efficient computational algorithms such as pathwise coordinate optimization \citep{FHHT08} and least angle regression \citep{EHJT04}.

One particular example is the combination of $L_1$ and $L_2$ regressions, in which $K = 2$, $\rho_1(t) = |t - b_0|$ and $\rho_2(t) = t^2$. Here $b_0$  denotes the median of error distribution $\varepsilon$.  If the
error distribution is symmetric, then $b_0 = 0$.  If the error
distribution is completely unknown, $b_0$ is unknown and  can be
estimated from the residual vector $\{\varepsilon_i^0\}$ or  being regarded as an additional
parameter and optimized  together with $\bbeta$ in
(\ref{eq8}). Another example is the combination of multiple quantile
check functions, that is,
$$
  \rho_k(t) = \tau_k (t-b_k)_+ + (1-\tau_k) (t-b_k)_-,
$$
where $\tau_k \in (0, 1)$ is a preselected quantile and $b_k$ is the
$\tau_k$-quantile of the error distribution.  Again, when $b_k$'s are
unknown, they can be estimated using the sample quantiles $\tau_k$ of the
estimated
residuals $\bvarepsilon^0$ or along with $\bbeta$ in
(\ref{eq8}). See  Section \ref{sec4} for additional discussion.

\section{Sampling properties and their applications}\label{sec3}

In this section,  we plan to establish the sampling properties of
estimator (\ref{eq5}) under the assumption that the number of
parameters (true dimensionality) $p$ and the number of non-vanishing
components (effective dimensionality) $s =  \| \bbeta^*\|_0$ satisfy $\log p = O(n^\delta)$ and $s = O(n^{\alpha_0})$ for some $\delta \in (0, 1)$ and $\alpha_0 \in (0, 1)$. Particular focus will be given to the oracle property of \cite{FL01}, but we will strengthen it and prove that estimator (\ref{eq5})  is an oracle estimator with overwhelming probability.  \cite{FL09} were among the first to discuss the oracle properties with NP dimensionality using
the full likelihood function in generalized linear models with a class
of folded concave penalties.  We work on a quasi-likelihood function and a class of weighted convex penalties.

\subsection{Asymptotic properties} \label{sec3.1}

To facilitate presentation, we relegate technical conditions and the details of  proofs to the Appendix.  We consider more generally the weighted $L_1$-penalized estimator with nonnegative weights $d_1, \cdots, d_p$.  Let
\begin{equation}\label{eq9}
 L_n(\bbeta)=\sum_{i=1}^n
\rho_{{\mathbf{w}}}\left( Y_i-\bX_{i}^T \bbeta \right) + n  \lambda_n \sum_{j=1}^p d_j \left|\beta_j\right|
\end{equation}
denote the penalized quasi-likelihood function.  The estimator in (\ref{eq5}) is a particular case of \eqref{eq9} and corresponds to the case with $d_j = \gamma_\lambda(|\beta_j^{(0)}|)/\lambda_n$.

Without loss of generality, assume that parameter $\bbeta^*$ can be arranged in
the form of $\bbeta^*=(\bbeta_1^{*}{}^T, \bzero^T)^T$, with $\bbeta_1^* \in R^s$ a vector of non-vanishing elements of $\bbeta^*$. Let us call $\hat{\bbeta}^{{\mathbf{o}}}=(\hat{\bbeta}_1^{\boldsymbol{\mathbf{o}} T},\mathbf{0}^T )^T \in R^p$ the biased oracle estimator, where $\hat{\bbeta}_1^{\mathbf{o}}$ is the minimizer of $L_n(\bbeta_1,\mathbf{0})$ in $R^s$ and $\mathbf{0}$ is the vector of all zeros in $R^{p-s}$. Here, we suppress the dependence of $\hat{\bbeta}^{\boldsymbol{\mathbf{o}}}$ on $\bw$ and $\bd = (d_1, \cdots, d_p)^T$.  The estimator $\hat{\bbeta}^{o}$ is called the biased oracle estimator, since the oracle knows the true submodel $\mathcal{M}_*=\{j:\beta_j^* \neq 0\}$, but nevertheless applies a penalized method to estimate the non-vanishing regression coefficients.  The bias becomes negligible when the weights in the first part are zero or uniformly small (see Theorem ~\ref{thm2}).
When the design matrix $\bS$ is non-degenerate, the function $L_n(\beta_1,\mathbf{0})$ is strictly convex and the biased oracle estimator is unique, where $\bS$ is a submatrix of $\bX$ such that $\bX=[\bS,\bQ]$ with $\bS$ and $\bQ$ being $n\times s$ and  $n \times (p-s)$ sub-matrices of $\bX$, respectively.

The following theorem shows that $\hat{\bbeta}^{\boldsymbol{\mathbf{o}}}$ is the unique minimizer of $L_n(\bbeta)$ on the whole space $\mathbf{R}^p$ with an overwhelming probability.
As a consequence, $\hat
\bbeta_\bw$  becomes the biased oracle. We establish the following theorem under conditions on
the non-stochastic vector $\bd$ (see Condition \ref{cond2}).  It is
also applicable to stochastic penalty weights as in \eqref{eq8}; see the remark following Condition~\ref{cond2}.

\begin{theorem}\label{thm1}
Under Conditions \ref{cond1}-\ref{cond4}, the estimators $\hat{\bbeta}^o$ and $\hat{\bbeta}_\bw$ exist and are unique on a set with probability tending to one.  Furthermore,
$$
P( \hat{\bbeta}_\bw =\hat{\bbeta}^{\boldsymbol{\mathbf{o}}} ) \geq 1 - (p-s)\exp\{- c n^{ (\alpha_0 - 2\alpha_1)_+  + 2 \alpha_2}\}
$$
for a positive constant $c$.
\end{theorem}

For the previous theorem to be nontrivial, we need to impose the
dimensionality restriction $\delta < (\alpha_0 - 2\alpha_1)_+ + 2
\alpha_2$, where $\alpha_1$ controls the rate of growth of the
correlation coefficients between the matrices $\bS$ and $\bQ$, the
important predictors and unimportant predictors (see
Condition~\ref{cond5}) and $\alpha_2 \in [0, 1/2)$ is a non-negative
constant, related to the maximum absolute value of the design matrix
[see Condition~\ref{cond4}].  It can be taken as zero and is
introduced to deal with the situation where $(\alpha_0 - 2
\alpha_1)_+$ is small or zero so that the result is trivial.  The
larger $\alpha_2$ is, the more stringent restriction is imposed on the choice of $\lambda_n$.
When the above conditions hold, the penalized composite
quasi-likelihood estimator $\hat{\bbeta}_\bw$ is equal to the biased
oracle estimator $\hat{\bbeta}^{\boldsymbol{\mathbf{o}}}$, with
probability tending to one exponentially fast.

\noindent{\bf Remark~{\color{red}4}:} The result of Theorem \ref{thm1} is stronger than the oracle property
defined in \cite{FL01} once the properties of $\hat{\bbeta}^o$ are
established (see Theorem~\ref{thm2}). It was formulated by
\cite{KCO08} for the SCAD estimator with polynomial dimensionality
$p$.  It implies not only the model selection consistency and but also sign consistency
\citep{ZY06,BRT08,BRT09}:
$$
   P(\mbox{sgn}(\hat{\bbeta}_\bw) =\mbox{sgn}({\bbeta}^{*}) )  = P(\mbox{sgn}(\hat{\bbeta}^o) =\mbox{sgn}({\bbeta}^{*}) )
   \to 1.
$$
In this way, the result of Theorem \ref{thm1} nicely unifies the two approaches in discussing the oracle property in high dimensional spaces.

Let $\widehat \bbeta_{\bw1}$ and $\widehat \bbeta_{\bw2}$ be  the first $s$ components and the remaining $p-s$ components of
$\widehat \bbeta_{\bw}$, respectively.  According to
Theorem~\ref{thm1}, we have $\widehat \bbeta_{\bw2} = \mathbf{0}$ with
probability tending to one. Hence, we only need to  establish the properties of $\widehat{\bbeta}_{\bw1}$.

\begin{theorem}\label{thm2}
Under Conditions \ref{cond1}-\ref{cond5}, the asymptotic bias of non-vanishing component $\widehat{\bbeta}_{\bw 1}$ is controlled by $D_n=\max \{d_j:j \in \mathcal{M}_*\}$ with
$$
    \| \widehat \bbeta_{\bw1} - \bbeta^{*}_1 \|_2 = O_P \left \{ \sqrt{s} ( \lambda_n D_n+ n^{-1/2}) \right \}.
$$
Furthermore, when $0 \leq \alpha_0 < 2/3$, $\widehat{\bbeta}_{\bw 1}$  possesses asymptotic normality:
\begin{equation}\label{eq10} 		
   \mathbf{b}^T (\bS^T \bS)^{1/2} \left(\widehat {\bbeta}_{\bw1} - \bbeta^{*}_1 \right)
    \stackrel{\mathcal{D}}{\to} \mathcal{N} \left(0, \sigma^2_{\bw} \right) 		 \end{equation}
where $\bb$ is a unit vector in $\mathbb{R}^s$ and
\begin{equation}\label{eq11}
\sigma^2_{\bw} = \frac{\sum_{k,l=1}^Kw_kw_lE[\psi_k(\varepsilon)\psi_l(\varepsilon)]}{\left(\sum_{k=1}^Kw_kE[\partial \psi_k(\varepsilon)]\right)^2}.
\end{equation}
\end{theorem}

Since the dimensionality $s$ depends on $n$, the asymptotic normality of
$\hat \bbeta_{\bw1}$ is not well defined in the conventional probability sense.  The arbitrary linear combination $\bb^T \hat \bbeta_{\bw1}$ is used to overcome the technical difficulty. In particular, any finite component of $\hat \bbeta_{\bw1}$ is asymptotically normal.
The result in Theorem~\ref{thm2} is also equivalent to the asymptotic normality of the linear combination $\bB^T \hat \bbeta_{\bw1}$ stated in \cite{FP04}, where $\bB$ is a $q\times s$ matrix, for any given finite number $q$.

This theorem relates to the results of \cite{P85} in classical setting (corresponding to $p=s$) where he established asymptotic normality of $M$-estimators  when the dimensionality is not higher than $o(n^{2/3})$. 

\subsection{Covariance Estimation}\label{sec3.2}

The asymptotic normality (\ref{eq10}) allows us to do statistical inference for non-vanishing components.  This requires an estimate of the asymptotic covariance matrix of $\widehat \bbeta_{\bw1}$.  Let $\hat \bvarepsilon = \bY - \bS^T \hat\bbeta_{\bw1}$ be the residual and $\hat{\varepsilon}_i$ be its $i$-th component.  A simple substitution estimator of $\sigma_{\bw}^2$ is
$$
\widehat \sigma_{\bw}^2 = \frac{n \sum_{k,l=1}^Kw_kw_l \sum_{i=1}^n \psi_k(\hat \varepsilon_i )\psi_l(\hat \varepsilon_i)}{\left(\sum_{k=1}^Kw_k  \sum_{i=1}^n \partial \psi_k(\hat \varepsilon_i )\right)^2}.
$$
See also the remark proceeding \eqref{eq7}.
Consequently, by \eqref{eq10}, the asymptotic variance-covariance matrix of $\widehat{\bbeta}_{\bw1}$ is given by
\begin{equation} \label{eq12}
     \widehat \sigma_{\bw}^2 (\bS^T \bS)^{-1}.
\end{equation}
Another possible estimator of the variance and covariance matrix is to apply the standard sandwich formula. In Section \ref{sec5}, through simulation studies, we show that this formula has good properties for both $p$ smaller and larger than $n$ (see Tables 3 and 4 and comments at the end of Section~\ref{sec5.1}).

\subsection{Choice of weights}\label{sec3.3}

Note that only the  factor $\sigma^2_{\bw}$ in equation \eqref{eq11} depends on the choice of $\mathbf{w}$ and it is invariant to the scaling of $\mathbf{w}$. Thus, the optimal choice of weights for maximizing efficiency of the estimator $\widehat \bbeta_{\bw1}$ is
\begin{eqnarray}\label{eq13}
\mathbf{w}_{opt} = \arg\!\min_{\mathbf{w}} \mathbf{w}^T\mathbf{M}\mathbf{w}  \ \ \ \ \ \mbox{s.t.} \ \ \ \mathbf{a}^T\mathbf{w}=1,  \ \ \  \mathbf{w} \geq 0
\end{eqnarray}
where $\bf M$ and $\ba$ are defined in Section~\ref{sec2} using an initial estimator, independent of the weighting scheme $\bw$.

\noindent{\bf Remark~{\color{red} 5}:} The quadratic optimization problem \eqref{eq13} does not have a closed form solution, but can  easily be solved numerically for a moderate $K$. The above efficiency gain, over the
least-squares, could  be better understood from the likelihood point of view. Let $f(t)$  denote the unknown error density. The most efficient loss function is the unknown log-likelihood function,
$-\log f(t)$. But since we have no knowledge of it, the set $\mathcal{F}_K$, consisting of convex combinations of $\{\rho_k(\cdot)\}_{k=1}^K$ given in \eqref{eq3},  could be viewed as a collection of basis functions used to approximate it. The broader the set $\mathcal{F}_K$ is, the
better it can approximate the log-likelihood function and the more efficient the estimator $\hat{\bbeta}^{a}$ in \eqref{eq8} becomes. Therefore, we refer to
$\rho_{\mathbf{w}}$  as the quasi-likelihood function.

\subsection{One-step penalized estimator} \label{sec3.4}

The restriction of $\mathbf{w} \geq 0$  guarantees the convexity of $\rho_{\mathbf{w}}$  so that the problem \eqref{eq5} becomes a convex optimization
problem.  However, this restriction may cause  substantial loss of efficiency in estimating $\hat{\bbeta}_{\bw 1}$ (see Table 1). We propose a one-step penalized estimator to  overcome this drawback while avoiding non-convex optimization.  Let $\hat{\bbeta}$ be the estimator based on the convex combination of loss functions \eqref{eq5} and $\hat{\bbeta}_1$ be its nonvanishing components. The one-step estimator is defined as
\begin{equation}\label{eq14}
\hat{\bbeta}^{\footnotesize{\mbox{os}}}_{\mathbf{w}1}=\hat\bbeta_{1} -
\left[\Omega_{n, {\mathbf{w}}}(\hat\bbeta_1)\right]^{-1} \Phi_{n,{\mathbf{w}}}(\hat\bbeta_1), \
\hat{\bbeta}^{\footnotesize{\mbox{os}}}_{\mathbf{w}2} =\mathbf{0},
\end{equation}
where $$\Phi_{n,{\mathbf{w}}}(\hat\bbeta_1)=\sum_{i=1}^n
\psi_{{\mathbf{w}}}(Y_i - \bS_i^T \hat\bbeta_{1}) \bS_i, $$
 $$\Omega_{n,{\mathbf{w}}}(\hat\bbeta_1)= \sum_{i=1}^n \partial \psi_{{\mathbf{w}}}(Y_i - \bS_i^T \hat\bbeta_1)\bS_i \bS_i^T.$$

\begin{theorem}\label{thm3}
Under Conditions \ref{cond1}-\ref{cond5},  if $\| \hat{\bbeta}_1 - \bbeta_1^*\| = O_p(\sqrt{s/n})$, then the one-step  estimator $\hat{\bbeta}^{\footnotesize{\mbox{os}}}_{\mathbf{w}}$ \eqref{eq14}  enjoys the asymptotic normality:
 \begin{equation} \label{eq15}
    \mathbf{b}^T (\bS^T \bS)^{1/2} \left(\hat {\bbeta}^{\footnotesize{\mbox{os}}}_{\mathbf{w}1} -\bbeta^{*}_1 \right) \stackrel{\mathcal{D}}{\to} \mathcal{N}
\left(0, \sigma^2_\mathbf{w} \right), 		 \end{equation}
provided that $s = o(n^{1/3})$, $\partial \psi(\cdot)$ is Lipchitz continous, and
$\lambda_{\max}( \sum_{i=1}^n \|\bS\|_i \bS_i \bS_i^T ) = O(n \sqrt{s})$,
where $\lambda_{\max}(\cdot)$ denote the  maximum eigenvalue of a matrix and $\sigma^2_\mathbf{w}$ are defined as in Theorem \ref{thm2}.
\end{theorem}

The one-step estimator \eqref{eq14} overcomes the convexity restriction  and  is always well defined, whereas \eqref{eq5} is not uniquely defined when convexity of
$\rho_{\mathbf{w}}$ is ruined. Note that if we remove the constraint of $w_k\geq 0$ ($k=1,...,K$), the optimal weight vector in \eqref{eq13} is equal to
$$
\mathbf{w}_{opt} =
\mathbf{M}^{-1}\mathbf{a} \ \mbox{ and } \ \sigma^2_{\mathbf{w}_{opt}}=(\mathbf{a}^T \mathbf{M}^{-1}\mathbf{a})^{-1}.
$$
This can be significantly smaller than the optimal variance obtained with convexity constraint, especially for multi-modal distributions (see Table 1).

The above discussion prompts a further improvement of the penalized adaptive composite quasi-likelihood in Section~\ref{sec2}.  Use \eqref{eq8} to compute the new residuals and new matrix $\bM$ and vector $\ba$.  Compute the optimal unconstrained weight $ \mathbf{w}_{opt} = \mathbf{M}^{-1}\mathbf{a}$ and the one-step estimator \eqref{eq14}.

\section{Examples}\label{sec4}

In this section, we discuss two  specific examples of penalized
quasi-likelihood regression. The proposed methods are complementary,
in the sense that the first one is computationally easy  but loses
some general flexibility while the second one is computationally
intensive  but efficient in a broader class of error distributions.

\subsection{Penalized Composite $L_1$-$L_2$ regression}\label{sec4.1}
First, we consider the combination of $L_1$ and $L_2$ loss functions, that is, $\rho_1(t)=|t-b_0|$ and $\rho_2(t)=t^2$. The nuisance parameter $b_0$ is  the median of the error
distribution.
Let $\hat{\bbeta}^{L_1-L_2}_{\bw} $ denote the corresponding penalized estimator as the solution to the minimization problem:
\begin{eqnarray}\label{eq17}
 \arg\!\min_{\boldsymbol{\beta},b_0} \   w_1 \sum_{i=1}^n  \left|Y_i - b_0- \bX_{i}^T\bbeta \right |+
 w_2  \sum_{i=1}^n\left(Y_i -  \bX_{i}^T \bbeta\right)^2  + n  \sum_{j=1}^p \gamma_{\lambda}(|\beta_j^{(0)}|)|\beta_j|.
\end{eqnarray}
If the error distribution is symmetric, then $b_0 =0$ and the minimization problem (\ref{eq17}) can be recast as a penalized weighted least square regression
\begin{equation*}
 \arg\!\min_{\boldsymbol{\beta}}  \sum_{i=1}^n  \left(\frac{w_1}{\left|Y_i- \bX_{i}^T\hat{\bbeta}^{(0)} \right |}+ w_2 \right) \left(Y_i -
\bX_{i}^T\bbeta\right)^2  + n  \sum_{j=1}^p \gamma_{\lambda}(|\beta_j^{(0)}|)|\beta_j|
\end{equation*}
which can be efficiently solved by pathwise coordinate optimization \citep{FHHT08} or least angle regression \citep{EHJT04}.

If $b_0 \ne 0$, the penalized least-squares problem (\ref{eq17}) is somewhat different from (\ref{eq5}) since we have an additional parameter $b_0$.   Using the same arguments, and treating $b_0$ as an additional parameter for which we solve in \eqref{eq17},  we can show that the conclusions of Theorems
\ref{thm2} and \ref{thm3} hold with the asymptotic variance equal to
\begin{equation}
\sigma^2_{L_1-L_2}(\mathbf{w})=  \frac{w_1^2/4 + w_2^2\sigma^2  + w_2w_1 B}{\left(w_1 f(b_0)+w_2 \right)^2}, \label{eq18}
\end{equation}
where $B = E[\varepsilon(I(\varepsilon>b_0)-I(\varepsilon<b_0))]$ and $f(\cdot)$ is the density of $\varepsilon$. This will hold when $b_0$ is either known or unknown.  Explicit optimization of \eqref{eq18} is not trivial and we go through it as follows.

Since $\sigma^2_{L_1-L_2}(\mathbf{w})$ is invariant to the scale of
$\mathbf{w}$, by setting $w_1/w_2=c\sigma$, we have
\begin{equation}\label{eq19}
\sigma^2_{L_1-L_2}(c)= \sigma^2 \frac{c^2/4 + 1  + a_\varepsilon c}{\left(b_\varepsilon c +1 \right)^2}.
\end{equation}
where $a_\varepsilon = B/\sigma$ and $b_\varepsilon = \sigma f(b_0)$.  Note that
$$
|B| \leq E |\varepsilon| [I(\varepsilon>b_0) + I(\varepsilon<b_0)] \leq \sigma.
$$
Hence, $|a_\varepsilon| \leq 1$ and
$
c^2/4 + 1  + a_\varepsilon c = (c/2 + a_\varepsilon)^2 + 1 - a_\varepsilon^2 \geq 0.
$

The optimal value of $c$ over $[0, \infty)$ can be easily computed. If $ a_\varepsilon b_\varepsilon  < 0.5$, then the optimal value is obtained at
\begin{equation} \label{eq20}
    c_{\varepsilon} = 2 (2 b_\varepsilon   - a_\varepsilon)_+ /(1 - 2 a_\varepsilon b_\varepsilon).
\end{equation}
In particular, when $2 b_\varepsilon \leq a_\varepsilon$, $c_{\varepsilon} = 0$, and the optimal choice is the least-squares estimator.  When $a_\varepsilon b_\varepsilon = 0.5$, if $2 b_\varepsilon \leq a_\varepsilon$, then the minimizer is $c_{\varepsilon} = 0$. In
all other cases, the minimizer is $c_\varepsilon =\infty$ i.e. we are left to use $L_1$ regression alone.

The above result shows the limitation of the convex combination, i.e. $c \geq 0$.  In many cases, we are left alone with the least-squares or least absolute deviation regression without improving
efficiency.  The efficiency can be gained and achieved by allowing negative weights via the one-step technique as in Section~\ref{sec3.4}.  Let $g(c) = (c^2/4 + 1  + a_\varepsilon c)/(b_\varepsilon c +1 )^2$.  The function $g(c)$ has a pole at $c = -1/b_\varepsilon$ and a unique critical point
\begin{equation} \label{eq21}
 c_{opt} = 2 (2 b_\varepsilon   - a_\varepsilon)/(1 - 2 a_\varepsilon b_\varepsilon),
\end{equation}
provided that $a_\varepsilon b_\varepsilon \not = 1/2$.  Consequently, the function $g(c)$ can not have any local maximizer (otherwise, from the local maximizer to the point $c = -1/b_\varepsilon$, there must exist a local minimizer, which is also a critical point). Hence, the minimum value is attained at $c_{opt}$.  In other words,
\begin{equation} \label{eq22}
\min_w \sigma^2_{L_1-L_2}(\mathbf{w}) = \sigma^2 \min_c g(c) = d_\varepsilon \sigma^2,
\end{equation}
where
\begin{equation}
    d_\varepsilon = g(c_{opt}) = (1-a_\varepsilon^2)/(4 b_\varepsilon^2-4a_\varepsilon b_\varepsilon+1).
\end{equation}
Since the denominator can be written as $(a_\varepsilon - 2 b_\varepsilon)^2 + (1 - a_\varepsilon^2)$, we have $d_\varepsilon \leq 1$, namely, it outperforms the least-squares estimator, unless $a_\varepsilon=2b_{\varepsilon}$.  Similarly, it can be shown that
$$
d_\varepsilon = \frac{1-a_\varepsilon^2}{4 b_\varepsilon^2 [ 1 - a_\varepsilon^2 + (2a_\varepsilon - 1/b_\varepsilon)^2/4]} \leq \frac{1}{4b_\varepsilon^2},
$$
namely, it outperforms the least absolute deviation estimation, unless $a_\varepsilon b_\varepsilon=1/2$.

When error distribution is symmetric unimodal, $b_\varepsilon \geq 1/\sqrt{12}$, according to Chapter 5 of
\cite{L83}.  The worst scenario for the $L_1$-regression in comparison with the $L_2$-regression  is the uniform distribution (see Chapter 5, \cite{L83}), which has the relatively efficiency of merely $1/3$. For such uniform  distribution, $a_\varepsilon = \sqrt{3}/2$ and $b_\varepsilon = 1/\sqrt{12}$, $d_\varepsilon = 3/4$, and $c_{opt} = -2/\sqrt{3}$.  Hence, the best $L_1$-$L_2$ is 4 times better than $L_1$ regression alone.  More comparisons about the weighted $L_1$-$L_2$ combination with $L_1$ and least-squares are given in Table 1(Section~\ref{sec4.3}).

\subsection{Penalized Composite Quantile Regression}\label{sec4.2}
The weighted composite quantile regression (CQR) was first studied by \cite{K84} in classical statistical inference setting. \cite{ZY08}  used equally weighted CQR (ECQR) for
penalized model selection with $p$ large but fixed. We show that the
efficiency  of ECQR can be substantially improved by properly
weighting and extend the work to the case of  $p \gg n$. Consider $K$ different quantiles,
$0<\tau_1<\tau_2<...<\tau_K<1$. Let $\rho_k(t)=\tau_k(t-b_k)_++(1-\tau_k)(t-b_k)_-$. The penalized composite quantile regression estimator $\hat{\bbeta}^{\footnotesize{\mbox{cqr}}}$ is defined  as
the solution to the minimization  problem
\begin{eqnarray} \label{eq24}
  \arg\!\min_{b_1,...,b_k,\bbeta} \  \sum_{k=1}^Kw_k\sum_{i=1}^n\rho_{k}\left(Y_i-\bX_{i}^T\bbeta \right) + n  \sum_{j=1}^p \gamma_{\lambda} (|\beta_j^{(0)}|) |\beta_j|,
\end{eqnarray}
where ${b}_k$ is the estimator of the nuisance parameter $b_k^*=F^{-1}(\tau_k)$, the $\tau_k$-th quantile of the error distribution.
Note that $b_1,\cdots, b_K$ are nuisance parameters and the minimization at \eqref{eq24} is done with respect to them too. After some algebra we can confirm that the conclusions of Theorems \ref{thm2} and \ref{thm3} continue to hold with the asymptotic variance as
\begin{equation} \label{eq25}
\sigma^2_{\footnotesize{\mbox{cqr}}}(\mathbf{w}) =  \frac{\sum_{k, k'=1}^K w_kw_{k'} ( \min(\tau_k, \tau_{k'}) -\tau_k\tau_{k'}) }{ \left(\sum_{k=1}^Kw_kf(F^{-1}(\tau_k))\right)^{2}}.
\end{equation}
As shown in \cite{K84} and \cite{B73}, when $K \to \infty$, the
optimally weighted CQR (WCQR) is as efficient as the maximum
likelihood estimator, always more efficient than
ECQR. Computationally, the minimization problem in equation (\ref{eq24}) can be casted as a large scale linear programming problem by expanding the covariate space with new  ancillary variables. Thus, it is computationally intensive to use too many quantiles. In Section \ref{sec4.3}, we can see that usually no more than ten quantiles are adequate for WCQR to approach the efficiency of MLE, whereas determining the optimal value of $K$ in ECQR seems difficult since the efficiency is not necessarily an increasing function of $K$ (Table 2). Also, some of the weights in $\mathbf{w}_{opt}$ are zero, hence making WCQR method  computationally less intensive than ECQR. From our experience in large $p$ and small $n$
situations, this reduction tends to be significant.

The optimal convex combination of quantile regression uses the weight
\begin{equation} \label{eq26}
\bw_{opt}^+ = \argmin_{\bw \geq 0, \ba^T \bw = 1} \bw^T \bM \bw,
\end{equation}
where $\ba = (f(F^{-1}(\tau_1)), \cdots, f(F^{-1}(\tau_K)))^T$ and
$\bM$ is a $K\times K$ matrix whose $(i, j)$-element is $\min(\tau_i,
\tau_{j}) -\tau_i\tau_{j}$.  The optimal combination of quantile
regression, which is obtained by using the one-step procedure, uses the
weight
\begin{equation} \label{eq27}
\bw_{opt} =  \bM ^{-1}\ba.
\end{equation}
Clearly, both combinations improve the efficiency of ECQR and the optimal combination is  most efficient  among the three (see Table 1).  When the error distributions are skewed or multimodal, the improvement can be substantial.

\subsection{Asymptotic Efficiency Comparison}\label{sec4.3}
In this section, we studied the asymptotic efficiency of proposed
estimators under several error distributions. For comparison, we also
included $L_1$ regression, $L_2$ regression and ECQR.
The error distribution ranges from
the symmetric to asymmetric distributions:    double
exponential (DE), $t$ distribution with degree of freedoms 4 ($t_4$),
normal $\mathcal{N}(0,1)$, Gamma $\Gamma (3,1)$,  Beta
$\mathcal{B}(3,5)$, a scale mixture of normals ($\MN_s$)
$0.1N(0,25)+0.9N(0,1)$ and a location mixture of normals ($\MN_l$)
$0.7 N(-1, 1) + 0.3 N(7/3, 1)$. To keep the comparison fair and to satisfy the first assumption of mean zero error terms, we first centered the error distribution to have mean zero.

Table 1 shows the asymptotic relative efficiency of each estimator
compared to MLE. $L_1$-$L_2^+$ and $L_1$-$L_2$ indicate
the optimal convex $L_1$-$L_2$ combination and optimal $L_1$-$L_2$
combination, respectively. While $L_1$ regression can have higher or
lower efficiency than  $L_2$ regression in different error
distributions,  $L_1$-$L_2^+$ and $L_1$-$L_2$ regressions are
consistently more efficient than both of them.
WCQR$^{+}$ denote the optimal convex combination of multiple
quantile regressions and WCQR represent the optimal combination.
In all quantile regressions, quantiles
$(\frac{1}{K+1},...,\frac{K}{K+1})$ were used. As shown in Table 1,
WCQR$^{+}$ and WCQR always outperform ECQR and the differences are more
significant in  double exponential distribution and asymmetric
distributions such as Gamma and Beta. In
DE, $t_4$ and $\mathcal{N}(0,1)$,  nine quantiles are
usually adequate for WCQR$^{+}$ and WCQR to achieve full
efficiency. In $\Gamma(3,1)$ and $\mathcal{B}(3,5)$, they need 29
quantiles to achieve efficiency close to MLE while the other
estimators are significantly inefficient. This difference is most expressed in multimodal distributions, MN$_s$ and MN$_l$, with WCQR outperforming all.
One of the possible problems with ECQR is that the efficiency does not necessarily increase with $K$, making the choice of $K$ harder.  For example, for the double exponential distribution, the relative efficiency decreases with $K$.  This is understandable, as $K=1$ is optimal:  Putting more and odd number of quantiles dilutes the weights.

\begin{table}
\caption{Asymptotic relative efficiency compared to MLE}
\centering
\begin{tabular}{lrccccccc}
\hline\hline
 $ f(\varepsilon)$&& DE& $t_4$ & $\mathcal{N}(0,1)$ & $\Gamma(3,1)$ & $\mathcal{B}(3,5)$ & $\MN_s$  & $\MN_l$ \\ \hline \hline
&$L_1$  &1.00&0.80&0.63& 0.29&0.41& 0.61&0.35 \\
&$L_2$  &0.50&0.35&1.00& 0.13&0.68&0.05&0.14\\
&$L_1$-$L_2^+$ &\textcolor{blue}{1.00}$^{\sharp}$&\textcolor{blue}{0.85}&\textcolor{blue}{1.00}&\textcolor{blue}{0.34}&\textcolor{blue}{0.68}& \textcolor{blue}{0.61}&\textcolor{blue}{0.63}\\
&$L_1$-$L_2$   &\textcolor{red}{1.00}$^{\natural}$&\textcolor{red}{0.85}&\textcolor{red}{1.00}&\textcolor{red}{0.44}&\textcolor{red}{0.80}&\textcolor{red}{0.61} &{\textcolor{red}{0.63}} \\
\hline
ECQR& $K=3$   & 0.84&0.94&0.86&0.43&0.59& 0.76&0.44 \\
&$5$  & 0.83&0.97&0.89&0.47&0.65& 0.78&0.50 \\
&$9$   & 0.82&0.97&0.92&0.49&0.68&0.77 &0.52 \\
&$19$  & 0.82&0.97&0.94&0.50&0.69& 0.75&0.54\\
&$29$  & \textcolor{green}{0.83}&\textcolor{green}{0.97}&\textcolor{green}{0.95}&\textcolor{green}{0.51}&\textcolor{green}{0.71}&\textcolor{green}{ 0.76}&\textcolor{green}{0.54}\\
\hline WCQR$^{+}$& $K=3$  & 0.95${^\dagger}$   &0.94&0.87&0.51&0.61& 0.76&0.60\\
&$5$& 0.96  &0.97&0.91&0.59&0.70&0.78 &0.69\\
&$9$ & 0.97  &0.98&0.95&0.69&0.78&0.79 &0.77\\
&$19$&0.98 &0.99&0.98&0.80&0.86& 0.80&0.83\\
&$29$& \textcolor{magenta}{0.99}&\textcolor{magenta}{0.99}&\textcolor{magenta}{0.99}&\textcolor{magenta}{0.85}&\textcolor{magenta}{0.90}& \textcolor{magenta}{0.80}&\textcolor{magenta}{0.84}\\
\hline WCQR& $K=3$  & 0.95${^\ddagger}$  &0.94&0.87&0.51&0.61& 0.76&0.61\\
&$5$& 0.96  &0.97&0.91&0.60&0.72&0.78 &0.76\\
&$9$ & 0.98 &0.98&0.95&0.70&0.80&0.79 &0.88\\
&$19$&0.99 &0.99&0.98&0.81&0.88& 0.92&0.95\\
&$29$& \textcolor{cyan}{0.99}
&\textcolor{cyan}{0.99}&\textcolor{cyan}{0.99}&\textcolor{cyan}{0.86}&\textcolor{cyan}{0.92}&
\textcolor{cyan}{0.93}&\textcolor{cyan}{0.97}\\
\hline
\end{tabular}
\end{table}

In Table 2 we illustrate both the adaptivity of the proposed composite QMLE methodology and computational efficiency of WCQR${^+}$ over
ECQR   by showing the positions of zero of  the optimal nonnegative weight vector $\mathbf{w}_{opt}^+$. For $K=9$,  only 1 quantile is needed  in the DE case, 5 and 6 quantiles are needed for MN$_l$ and MN$_s$ and 7 quantiles for  $t_4$, Gamma and Beta. Only in the  normal distribution,
all 9 quantiles are used. Therefore, WCQR${^+}$ can dramatically reduce the computational complexity of ECQR in large scale optimization problems where $p\gg n$.

\begin{table}
\caption{Optimal weights of convex composite  quantile regression with $K=9$ quantiles} \label{tab2}
\centering
\begin{tabular}{rrrrrrrrrr}
  \hline \hline
&$f(\varepsilon)$& DE & $t_4$ & $\mathcal{N}(0,1)$ & $\Gamma$(3,1) & $\mathcal{B}$(3,5)& $\MN_s$  & $\MN_l$& \\
  \hline \hline
&Quantile\\
&  1/10 & 0 		& 0 		& 0.20 	& 0.56 	& 0.39 		& 0.06 &0.36& \\
&  2/10 & 0 		& 0.12 	& 0.11	& 0.15 	& 0.10		& 0.23&0.11&\\
&  3/10 & 0 		& 0.14 	& 0.09 	& 0.08	 & 0.11		& 0.17&0.10 &\\
&  4/10 & 0 		& 0.14 	& 0.08 	& 0.06 	& 0.05	 	& 0.10& 0.01&\\
&  5/10 & 1	 	& 0.16 	& 0.06 	& 0.05 	& 0 		 	& 0.14&0&\\
&  6/10 & 0	 	& 0.14 	& 0.08 	& 0.04	& 0		 	& 0& 0&\\
& 7/10 & 0 		& 0.14	& 0.09 	& 0 		& 0.05	 	& 0&0& \\
&  8/10 & 0 		& 0.12 	& 0.11 	& 0	 	& 0.09 		& 0 &0&\\
&  9/10 & 0 		& 0  		& 0.20 	& 0.05	 & 0.20 		& 0.30&0.29& \\
   \hline
\end{tabular}
\end{table}

\section{Finite Sample Study}\label{sec5}

\subsection{Simulated example}\label{sec5.1}

In the simulation study, we consider the classical linear model for testing variable selection methods used by \cite{FL01}
\[
y=\mathbf{x}^T \boldsymbol{\beta^*} +  \varepsilon, \ \ \mathbf{x} \sim N(0, \mathbf{\Sigma}_{\mathbf{x}}), \ \ (\mathbf{\Sigma}_{\mathbf{x}})_{i,j}=(0.5)^{|i-j|}.
\]
 The error vector varies from uni- to multi-modal and heavy to light tails distributions in the same way as in  Tables 1 and 2,  and is centered to have mean zero. The data has  $n=100$ observations. We considered two settings where $p=12$ and $p=500$, respectively. In both settings, $(\beta_1,\beta_2,\beta_5)= (3,1.5,2)$ and the other coefficients are equal to
 zero.
We implemented penalized $L_1$, $L_2$, composite $L_1$-$L_2^{+}$, $L_1$-$L_2$,
ECQR, WCQR$^+$ and WCQR
using quantiles
$(10\%, 20\%,\cdots, 90\%)$. The local linear approximation of SCAD penalty \eqref{eq6} was used and
the tuning parameter in the penalty was selected using five fold cross validation. We compared different methods by: (1) model error, which is defined as
$\mbox{ME}(\widehat{\bbeta})=(\widehat{\bbeta}- \bbeta^*)^T E(\boldsymbol{X}^T\boldsymbol{X})(\widehat{\bbeta}- \bbeta^*)$;  (2)  the number of
correctly classified non-zero coefficients , i.e. the true positive
(TP);  (3) the number of incorrectly classified zero coefficients,
i.e. the false positive (FP); (4) the multiplier $\hat{\sigma}_\bw$ of
the standard error (SE)\eqref{eq12}. A total of 100 replications were
performed and the median of model error (MME), the
average of TP and FP are reported in Table 3. The median model errors of oracle estimators were  calculated as the benchmark for comparison.
\begin{tiny}
\begin{table}
\caption{Simulation results ($n=100,p=12$) where $\dagger$, $\ddagger$  represent Median model error (MME) of the oracle  and  penalized estimator respectively}\label{tab3}
\centering
\begin{tabular}{lr|ccccccc}
\hline \hline $f(\varepsilon)$&&DE & $t_4$  & $\mathcal{N}(0,3)$&$\Gamma$(3,1) & $\mathcal{B}$(3,5)& MN$_s$ \\ \hline \hline
 $L_1$ & Oracle  & 0.029$^{\dagger}$  &0.050 &0.122 &0.082 &0.0010 & 0.043  \\
 & {Penalized}  & 0.035$^\ddagger$&0.053&0.128 &0.097 & 0.0011&0.051 \\
& (TP, FP)  & (3,1.83) & (3,0.8) &(3,0.84) & (3,1)  &(3,0.54) &(3,0.93) \\
&SD$\times10^{2}$&0.646 &0.767&0.570&0.950 &0.112&0.244
 \\
\hline
 $L_2$&Oracle&   0.047  & 0.043 &0.073 &0.064&0.00056  &0.083 \\
 & {Penalized} &  0.059  & 0.054 &0.106 &0.100 &0.0011& 0.091 \\
 &  (TP, FP) & (3,0.82) & (3,1.61) &(3,1.89)&(3,1.35)&(3,3.76) & (3,1.47)\\
 &SD$\times10^{2}$& 0.779&0.762&0.485&0.869&0.129&0.179
 \\
\hline
 $L_1$-$L_2^+$& Oracle   &  0.036& 0.043 &0.070 &0.070& 0.00061& 0.051\\
 & {Penalized}  & 0.037 & 0.049 & 0.102&0.099&0.00077 & 0.058 \\
 &(TP, FP) & (3,2.49) &  (3,2.39) & (3,1.97) &(3,2.09) & (3,2.42) & (3,2.69)\\
 &SD$\times10^{1}$& 0.717&0.702&0.518&0.876&0.095&0.169
 \\
\hline
 $L_1$-$L_2$ & Oracle  & 0.036  & 0.043 & 0.070 &0.063&0.00060 & 0.051\\
 & {Penalized} &  0.037& 0.049 & 0.102 & 0.078& 0.00063& 0.058 \\
  &(TP, FP) & (3,2.49)& (3,2.39) & (3,1.97) &(3,2.05) &(3,2.42) &(3,2.69)\\
 &SD$\times10^{2}$&0.717&0.702&0.518&0.846&0.075&0.169
 \\
\hline
 ECQR & Oracle & 0.031 & 0.046 & 0.069  &0.063 & 0.00065 & 0.033 \\
 & {Penalized} &0.042  & 0.046  & 0.107 & 0.074&0.00091 & 0.040 \\
& (TP, FP)  & (3,1.88) & (3,1.57) & (3,2.04) &(3,1.83)& (3,1.88) &(3,1.38)  \\
&SD$\times10^{2}$& 0.654&0.562&0.488&0.813&0.087&0.177
 \\\hline
 WCQR$^+$& Oracle & 0.033& 0.047 & 0.068&0.052&0.00065 & 0.036\\
& {Penalized}&  0.039 &0.041 & 0.100&0.054& 0.00070& 0.037\\
& (TP, FP)  &(3,0.55) & (3,1.47) & (3, 0.74)& (3,0.61)&(3,0.98)& (3,0.62)\\
&SD$\times10^{1}$&0.440&0.612&0.498&0.715&0.071&0.174
 \\ \hline
WCQR & Oracle & 0.033& 0.047& 0.068&0.048& 0.00058 & 0.028\\
& {Penalized}&  0.039 & 0.041&  0.100 & 0.050&0.00062 & 0.030\\
& (TP, FP)  &(3,0.55) & (3,1.47) & (3, 0.74)&(3,0.61)&(3,0.98) & (3,0.62)\\
&SD$\times10^{1}$&0.440 &0.612 &0.498&0.650&0.061&0.132
 \\ \hline
\end{tabular}
\end{table}
\end{tiny}

From the results presented in Table 3 and Table 4, we can see that
penalized composite $L_1$-$L_2^+$ regression takes the smaller of the two model errors of $L_1$ and $L_2$ in all distributions  except in
$\mathcal{B}(3,5)$  where it outperforms both.  As expected,  optimal $L_1$-$L_2$ outperforms  $L_1$-$L_2^+$ and brings a smaller number of FP, especially in multimodal and unsymmetric distributions. Also, both $L_1$-$L_2^+$ and $L_1$-$L_2$  perform
reasonably well  when compared to ECQR, but with much less
computational burden.  WCQR$^{+}$ and WCQR in both
Tables 3 and 4 have smaller model errors  and smaller number of
false positives than ECQR.
Similar conclusions can be made from Figure 1, which compares the boxplots of the model errors of the five methods (WCQR$^{+}$ and $L_1-L_2^{+}$ are not shown) under
different distributions in the case of $n=100,p=500$.  For $p \ll n$ in Table 3 we didn't include LASSO estimator since it behaves reasonably well in that setting. For $p \gg n$ in Table 4, we included
LASSO estimator as a reference. Table 4 shows that LASSO has bigger
model errors, more false positives and higher standard errors (usually by a factor of $10$) than any other five SCAD based methods discussed.

In addition to the ME in Tables 3 and 4, we reported the multiplier $\hat\sigma_{\mathbf{w}}$ of the asymptotic variance (see equation \eqref{eq12}). Being the only part  of SE that depends on the choice of weights $\bw$ and loss functions $\rho_k$, we explored it's behavior when the dimensionality grows from $ p\ll n$ to $p\gg n$. Both Tables 3 and 4 confirm the stability of the formula throughout the two settings and all five CQMLE methods. Only Lasso estimator being unable to specify the correct sparsity set when $p \gg n$, inflates $\hat\sigma_{\mathbf{w}}$ for one order of magnitude compared to other CQMLEs. Note that WCQR$^+$ keeps the smallest value of $\hat\sigma_{\mathbf{w}}$ and all $L_1$-$L_2$,$L_1$-$L_2^+$, WCQR and WCQR$^+$ have smaller SEs than the classical $L_1$, $L_2$ or ECQR methods.

\begin{tiny}
\begin{table}
\caption{Simulation results ($n=100,p=500$) were  $\dagger$, $\ddagger$ are Median model error (MME) of  oracle   and penalized estimator respectively }\label{tab4}
\centering
\begin{tabular}{lr|cccccccccc}
\hline \hline  $f(\varepsilon)$& &DE&  $t_4$ & $\mathcal{N}$(0,3)  & $\Gamma$(3,1) & $\mathcal{B}$(3,5) &MN$_s$ \\ \hline \hline
Lasso&Oracle &0.039$^\dagger$ &  0.039 & 0.035 & 0.0719 &0.062& 0.176 \\
 &Penalized &1.775$^\ddagger$ & 1.759&8.687  & 2.662  &1.808&6.497 \\
&(TP,FP)& (3,94.46)&  (3,94.26)& (3,96.80)  & (3,95.59)&(3,86.88)&(3,96.55) \\
 &SD$\times 10^{2}$& 3.336&3.257&0.578 &3.167& 0.989&0.539 \\
 \hline
  $L_1$ &Oracle  & 0.025 &0.031 &0.382 &  0.096&0.0094& 0.281\\
 &Penalized &0.035  &0.039 & 1.342  &0.131  &0.0120 &  0.514\\
 &(TP,FP)& (3,4.53) &(3,4.47) & (3,5.32) & (3,4.56) & (3,8.10)& (3,4.58)  \\
 &SD$\times 10^{2}$&0.268&0.274&0.144&0.461&0.215 &0.101\\
\hline
$L_2$&Oracle &0.035&  0.043  &0.207 &0.078& 0.0057 &0.187 \\
 &Penalized &0.093 &0.086  & 1.187 &0.175 &0.0073&0.764\\
 &(TP,FP)& (3,12.31)  &  (3,10.64) &(3,11.00)& (3,8.02)& (3,18.75) &(3,16.93)\\
  &SD$\times 10^{1}$&0.865  &0.828 &0.281&0.168&0.396&0.238\\
\hline
$L_1$-$L_2^+$ &Oracle&0.193 & 0.035 & 0.224  & 0.080  &0.0061&0.195 \\
 &Penalized& 0.036 & 0.036 &1.160  & 0.097 &0.0077&0.576\\
& (TP,FP)& (3,17.92) &(3,12.58) &  (3,15.87)  & (3,15.43) & (3,14.05)& (3,17.92) \\
 & SD$\times 10^{2}$& 0.226 &0.235  &0.396&0.144 &0.235 &0.207\\
\hline
$L_1$-$L_2$ &Oracle&0.035 & 0.035  & 0.224& 0.079 &0.0050&0.195  \\
 &Penalized&0.036 &0.036  & 1.160  & 0.095 & 0.0069 &0.576 \\
&(TP,FP)& (3,17.92) &(3,12.58)  & (3,15.87)  & (3,15.43) & (3,14.05) & (3,17.92)\\
&SD$\times 10^{2}$&0.226& 0.235  &0.905 &0.150&0.190&0.207\\ \hline
ECQR &Oracle& 0.029  &0.024  & 0.252  & 0.057 & 0.0064  & 0.207\\
&Penalized& 0.060 &0.070  & 0.764  & 0.148    &0.0118 & 0.599
\\ &(TP,FP)& (3,8.71)&(3,8.43)& (3,7.78) &(3,9.59) & (3,9.69) &(3,8.91) \\
 &SD$\times 10^{1}$&0.469 &0.475  &0.153 &0.716&0.213& 0.139\\ \hline
 WCQR$^+$&Oracle&0.028 & 0.027& 0.223  &  0.050 & 0.0066 & 0.204\\
&Penalized&0.045 &0.037& 0.595   &  0.079  &0.0076& 0.368
\\ &(TP,FP)& (3,3.97)&(3,3.76) &(3, 3.93)   &(3,3.66)&(3,4.85)& (3,4.05) \\
 &SD$\times 10^{1}$&0.244 &0.266 &0.112  &0.273&0.120& 0.084\\ \hline
 WCQR&Oracle&0.028& 0.027& 0.223 & 0.048 &0.0048 &0.160\\
&Penalized&0.045&  0.037 &0.595 & 0.062 &0.0060& 0.280
\\ &(TP,FP)& (3,3.97)&(3,3.76)& (3,3.93) & (3,3.66)&(3,4.85)&(3,4.05)\\
 &SD$\times 10^{1}$&0.224&0.219 &0.112  &0.180&0.110& 0.060 \\ \hline
\end{tabular}
\end{table}
\end{tiny}

\begin{figure}\label{fig:boxplot}
\centering
\includegraphics[width=\textwidth]{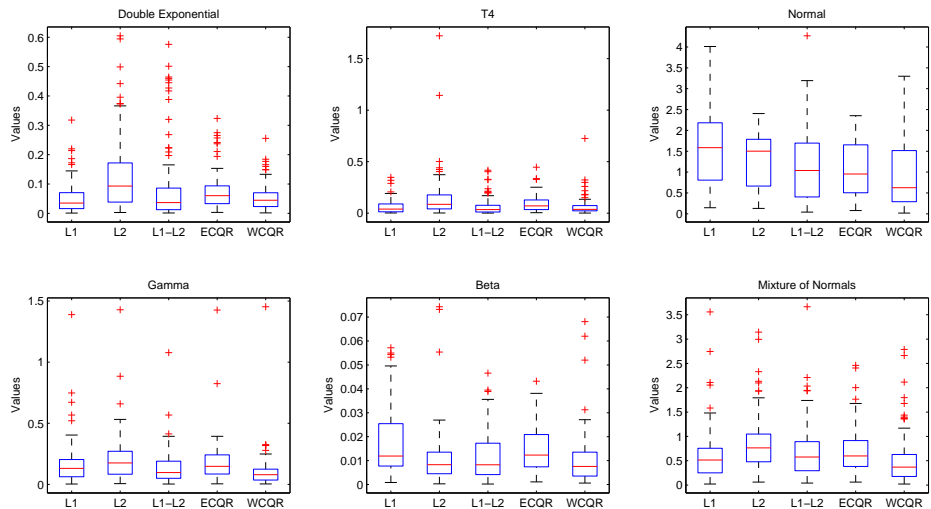}
\caption{Boxplots of Median model error (MME) of $L_1$, $L_2$, $L_1$-$L_2$, ECQR and  WCQR methods under different distributional settings with  $n=100$,$p=500$}
\end{figure}

\subsection{Real Data Example}\label{sec5.2}
In this section, we applied proposed methods to  expression quantitative trait locus (eQTL) mapping. Variations  in  gene expression levels may be related to  phenotypic variations such as
susceptibility to diseases and response to drugs. Therefore, to understand the genetic basis of  gene expression, variation  is an important topic in genetics. The availability of genome-wide single
nucleotide polymorphism (SNP) measurement has made it possible and reasonable to perform the high resolution eQTL mapping on the scale of nucleotides. In our analysis, we conducted the {\it cis}-eQTL
mapping  for the gene CCT8. This gene is located within the Down Syndrome Critical Region on human chromosome 21, on the minus strand. The over expression of CCT8 may be associated with Down syndrome
phenotypes.

We used the SNP genotype data and  gene expression data for the 210 unrelated individuals of the International HapMap project (International HapMap Consortium, 2003) \nocite{citeulike:587199}, which
include 45 Japanese in Tokyo, Japan, 45 Han Chinese in Beijing, China,  60 Utah parents with ancestry from northern and western Europe (CEPH) and 60 Yoruba parents in Ibadan, Nigeria and they are
available in PLINK format (Purcell, et al 2007) \nocite{Purcell_et_al07} [http://pngu.mgh.harvard.edu/purcell/plink/]. We included in the  analysis more than 2 million SNPs with minor allele frequency
greater than 1\% and missing data rate less than 5\%. The gene expression data were generated by Illumina Sentrix Human-6 Expression BeadChip and have been normalized (ith quantile normalization
across replicates and median normalization across individuals) independently for each population (Stranger, et al 2007) \nocite{citeulike:1097231} [ftp://ftp.sanger.ac.uk/pub/genevar/].

 Specifically, we considered the {\it cis}-candidate region to start 1 Mb upstream of the transcription start site (TSS) of CCT8 and to end 1 Mb downstream of the transcription end site (TES), which
 includes 1955 SNPs in Japanese and Chinese, 1978 SNPs in CEPH and 2146 SNPs in Yoruba.
 In the following analysis, we grouped Japanese and Chinese together into the Asian population and analyzed the three populations Asian, CEPH and Yoruba separately. The additive coding of SNPs (e.g.
 0,1,2) was adopted and was treated as categorical variables instead
 of continuous ones to allow non-additive effects, i.e., two dummy
 variables will be created for categories 1 and 2 respectively.  The category 0 represents the major, normal population.
The missing SNP measurements were imputed as 0's. The response variable is the gene expression level of gene CCT8, measured by microarray.

In the first step, the ANOVA F-statistic was computed for each SNP independently and a version of independent screening method of  \cite{FL08} was implemented. This method is  particularly
computationally efficient in ultra-high dimensional problems and here we retained the top 100 SNPs with the largest F-statistics. In the second step, we  applied to the screened data the penalized
$L_2$, $L_1$, $L_1$-$L_2^+$, $L_1$-$L_2$, ECQR, WCQR$^+$  and WCQR with local linear
approximation of SCAD penaly. All the four composite quantile regressions used quantiles at
$(10\%,...,90\%)$. LASSO was used as the initial estimator and the
tuning parameter in both LASSO and SCAD penalty was chosen by five
fold cross validation. In all the three populations, the $L_1$-$L_2$ and $L_1$-$L_2^+$ regressions reduced to $L_2$ regression. This is not unexpected due to the gene expression normalization procedure. In addition, WCQR reduced to WCQR$^+$.
 The selected SNPs, their coefficients and distances from transcription
 starting site (TSS) are summarized in Tables 5, 6 and 7.

In Asian population (Table 5), the five methods are reasonably
consistent in not only  variables selection but also  coefficients
estimation (in terms of signs and order of magnitude). WCQR uses the weights $(0.19,0.11,0.02,0,0.12,0.09,0.18,$ $0.19,0.10)$.
There are four SNPs  chosen by all the five methods. Two of them, rs2832159 and rs2245431, up-regulate gene expression while rs9981984 and rs16981663 down-regulate gene expression. The ECQR selects
the largest set of SNPs while $L_1$ regression selects the smallest set.
\begin{table}
\caption{eQTLs for gene CCT8 in Japanese and Chinese ($n=90$). ** is the indicator for SNP equal to 2 and otherwise is the indicator for 1. SE of the estimates is reported in the parenthesis.}\label{tab:jpt}
\centering
\begin{tabular}{rlrrrrrr}
  \hline \hline
 & SNP & $L_2$ & $L_1$-$L_2^+$ & $L_1$ & ECQR & WCQR$^+$ & Distance from\\
&&&$L_1$-$L_2$ &&& WCQR &  TSS (kb)\\
  \hline
  \hline
 & rs16981663 & -0.11 \scriptsize(0.03) & -0.11 \scriptsize(0.03) & -0.09 \scriptsize(0.04) & -0.10  \scriptsize(0.03)& -0.09 \scriptsize(0.03) & -998 \\
 & rs16981663$^{**}$ & 0.08 \scriptsize(0.06)& 0.08 \scriptsize(0.06)&  & 0.04 \scriptsize(0.06) &  & -998 \\
 & rs9981984 & -0.12 \scriptsize(0.03)& -0.12 \scriptsize(0.03)& -0.10 \scriptsize(0.04)& -0.09 \scriptsize(0.03)& -0.12 \scriptsize(0.03)& -950 \\
 & rs7282280 &  &  &  & 0.05 \scriptsize(0.03)&  & -231 \\
 & rs7282280$^{**}$ &  &  &  & -0.07 \scriptsize(0.05)&  & -231 \\
 & rs2245431$^{**}$ & 0.33 \scriptsize(0.10)& 0.33 \scriptsize(0.10)& 0.36 \scriptsize(0.11) & 0.37 \scriptsize(0.09) & 0.38 \scriptsize(0.10)& -89 \\
 & rs2832159 & 0.21 \scriptsize(0.04)& 0.21 \scriptsize(0.04) & 0.30 \scriptsize(0.04)& 0.20 \scriptsize(0.04)& 0.23 \scriptsize(0.04)& 13 \\
 & rs1999321$^{**}$ & 0.11 \scriptsize(0.07)& 0.11 \scriptsize(0.07)&  & 0.14 \scriptsize(0.07)&  & 84 \\
 & rs2832224 & 0.07 \scriptsize(0.03)& 0.07 \scriptsize(0.03)&  & 0.06 \scriptsize(0.03)& 0.04 \scriptsize(0.03)& 86 \\
  \hline
\end{tabular}
\end{table}
In CEPH population (Table 6), the five methods consistently selected
the same seven SNPs with only ECQR choosing two additional SNPs. WCQR
uses the weight $(0.19,0.21,0,0.04,$ $0.03,0.07,0.1,0.21,0.15)$. The coefficient estimations were also highly
consistent. Deutsch et al (2007)\nocite{citeulike:765404} performed a similar {\it cis}-eQTL mapping for the gene CCT8 using the same CEPH data as here. They considered a 100kb region surrounding the
gene, which contains 41 SNPs. Using ANOVA with correction for multiple tests, they identified four eQTLs, rs965951, rs2832159, rs8133819 and rs2832160, among which rs965951 possessing the smallest
p-value. Our analysis verified rs965951 to be an eQTL but did not find the other SNPs to be associated with the gene expression of CCT8. In other words, conditioning on the presence of SNP rs965951 the other three make little additional contributions.
\begin{table}
\caption{eQTLs for gene CCT8 in CEPH ($n=60$). ** is the indicator for SNP equal to 2 and otherwise is the indicator for 1. SE of the estimates is reported in the parenthesis.}\label{tab:ceu}
\centering
\begin{tabular}{rlrrrrrr}
  \hline \hline
 & SNP & $L_2$ & $L_1$-$L_2^+$ & $L_1$ & ECQR & WCQR$^+$ & Distance from\\
&&&$L_1$-$L_2$ &&& WCQR &  TSS (kb)\\
  \hline
  \hline
  & rs2831459 &0.20 \scriptsize (0.07)&0.20 \scriptsize (0.07)&0.19 \scriptsize (0.08)&0.17 \scriptsize (0.07)&0.18 \scriptsize (0.07)&-999 \\
  & rs7277536 &0.18 \scriptsize(0.09)&0.18 \scriptsize(0.09)&0.09 \scriptsize(0.11)&0.14 \scriptsize(0.09)&0.23 \scriptsize(0.09)&-672 \\
  & rs7278456$^{**}$ &0.36 \scriptsize(0.11)&0.36 \scriptsize(0.11)&0.21 \scriptsize(0.13)&0.40 \scriptsize(0.11)&0.35 \scriptsize(0.11)&-663 \\
  & rs2248610 &0.08 \scriptsize(0.04)&0.08 \scriptsize(0.04)&0.09 \scriptsize(0.05)&0.10 \scriptsize(0.05)&0.06 \scriptsize(0.05)&-169 \\
  & rs965951 &0.11 \scriptsize(0.05)&0.11 \scriptsize(0.05)&0.13 \scriptsize(0.06)&0.03 \scriptsize(0.06)&0.12 \scriptsize(0.05)&-13 \\
  & rs3787662 &0.12 \scriptsize(0.06)&0.12 \scriptsize(0.06)&0.08 \scriptsize(0.07)&0.13 \scriptsize(0.06)&0.12 \scriptsize(0.06)&78 \\
  & rs2832253 &&&&0.10 \scriptsize(0.07)&&117 \\
  & rs2832332 &&&&0.08 \scriptsize(0.05)&&382 \\
  & rs13046799 &-0.16 \scriptsize (0.05)&-0.16 \scriptsize(0.05)&-0.14 \scriptsize(0.06)&-0.14 \scriptsize(0.05)&-0.16 \scriptsize(0.05)&993 \\
   \hline
\end{tabular}
\end{table}
The analysis of Yoruba population yields a large number of eQTLs
(Table 7). The ECQR again selects the largest set of 44 eQTLs. The
$L_1$ regression selects 38 eQTLs. The $L_2$ regression and WCQR both
select 27 SNPs, 26 of which are the same. WCQR uses the weight
$(0.1,0,0.17,0.16,0.11,0.3,0,0,0.16)$. The coefficients estimated by different methods are mostly consistent (in terms of signs and order of magnitude), except that the coefficients estimates for rs8134601, rs7281691, rs6516887 and rs2832159 by ECQR and $L_1$ have different signs from those of $L_2$ and WCQR.
\begin{table}
\caption{eQTLs of gene CCT8 in Yoruba $(n=60)$;  ** is the indicator for SNP equal to 2 and otherwise is the indicator for 1. SE of the estimates is reported in the parenthesis.}\label{tab:yri}
\centering
\begin{tabular}{rlrrrrrc}
  \hline \hline
 & SNP & $L_2$ & $L_1$-$L_2^+$ & $L_1$ & ECQR & WCQR$^+$ & Distance from\\
&&&$L_1$-$L_2$ &&& WCQR &  TSS (kb)\\
  \hline
  \hline
 & rs9982023$^{**}$ &  &  & 0.12 \scriptsize(0.05)& 0.14 \scriptsize(0.04)&  & -531 \\
 & rs1236427 &  &  &  & 0.15 \scriptsize(0.04)&  & -444 \\
 & rs2831972 & -0.22 \scriptsize(0.06)& -0.22 \scriptsize(0.06)& -0.16 \scriptsize(0.07)& -0.30 \scriptsize(0.05)& -0.30 \scriptsize(0.06)& -360 \\
 & rs2091966$^{**}$ & -0.21 \scriptsize(0.11)& -0.21 \scriptsize(0.11)& -0.57 \scriptsize(0.16)& -0.39 \scriptsize(0.13)& -0.20 \scriptsize(0.11)& -358 \\
 & rs2832010 & -0.04 \scriptsize(0.03)& -0.04 \scriptsize(0.03)& -0.18 \scriptsize(0.08)& -0.32 \scriptsize(0.05)& -0.07 \scriptsize(0.03)& -336 \\
 & rs2832024 &  &  & 0.14 \scriptsize(0.09)& 0.26 \scriptsize(0.06)&  & -332 \\
 & rs2205413 & -0.08 \scriptsize(0.04)& -0.08 \scriptsize(0.04)& -0.15 \scriptsize(0.05)& -0.16 \scriptsize(0.04) & -0.04 \scriptsize(0.03)& -330 \\
 & rs2205413$^{**}$&  &  &  & -0.29 \scriptsize(0.05)&  & -330 \\
 & rs2832042$^{**}$ & 0.14 \scriptsize(0.04)& 0.14 \scriptsize(0.04)& 0.23 \scriptsize(0.05)& 0.23 \scriptsize(0.04)& 0.13 \scriptsize(0.04)& -330 \\
 & rs2832053$^{**}$ &  &  &  & -0.12 \scriptsize(0.13)&  & -315 \\
 & rs2832053 &  & & 0.09 \scriptsize(0.04)& 0.06 \scriptsize(0.02)&  & -315 \\
 & rs8130766 & -0.01 \scriptsize(0.03)& -0.01 \scriptsize(0.03)& -0.14 \scriptsize(0.05)& -0.10 \scriptsize(0.03)& -0.04 \scriptsize(0.03)& -296 \\
 & rs16983288$^{**}$ & -0.13 \scriptsize(0.07)& -0.13 \scriptsize(0.07)& -0.28 \scriptsize(0.08)& -0.28 \scriptsize(0.05)& -0.15 \scriptsize(0.06)& -288 \\
 & rs16983303 & -0.06 \scriptsize(0.03)& -0.06 \scriptsize(0.03)& -0.10 \scriptsize(0.02)& -0.15 \scriptsize(0.03)& -0.09 \scriptsize(0.03)& -283 \\
 & rs8134601$^{**}$ & 0.18 \scriptsize(0.11)& 0.18 \scriptsize(0.11)& 0.15 \scriptsize(0.12)& 0.16 \scriptsize(0.07)& 0.19 \scriptsize(0.10) & -266 \\
 & rs8134601 & -0.16 \scriptsize(0.12)& -0.16 \scriptsize(0.12)& 0.08 \scriptsize(0.15)& 0.25 \scriptsize(0.11)& -0.17 \scriptsize(0.11)& -266 \\
 & rs7276141$^{**}$ &  &  & -0.06 \scriptsize(0.12)& 0.15 \scriptsize(0.11)& & -264 \\
 & rs7281691 & 0.23 \scriptsize(0.10)& 0.23 \scriptsize(0.10)& -0.03 \scriptsize(0.13)& -0.18 \scriptsize(0.10)& 0.26 \scriptsize(0.09)& -263 \\
 & rs7281691$^{**}$ & -0.14 \scriptsize(0.09)& -0.14 \scriptsize(0.09)& -0.05 \scriptsize(0.13)& -0.23 \scriptsize(0.09)& -0.12 \scriptsize(0.09)& -263 \\
 & rs1006903$^{**}$ & -0.01 \scriptsize(0.05)& -0.01\scriptsize(0.05)& 0.13 \scriptsize(0.06)& 0.07 \scriptsize(0.04)& 0.01 \scriptsize(0.05)& -246 \\
 & rs7277685 & & & 0.07 \scriptsize(0.05)& 0.06 \scriptsize(0.03)&  & -240 \\
 & rs9982426 & 0.02 \scriptsize(0.03)& 0.02 \scriptsize(0.03)& 0.12 \scriptsize(0.04)& 0.18 \scriptsize(0.05)&  & -238 \\
 & rs2832115 &  &  &  & -0.08 \scriptsize(0.05)& & -225 \\
 & rs11910981 & -0.09 \scriptsize(0.03)& -0.09 \scriptsize(0.03)& -0.15 \scriptsize(0.03)& -0.19 \scriptsize(0.03)& -0.08 \scriptsize(0.03)& -160 \\
 & rs2243503 &  &  &  & 0.07 \scriptsize(0.06) &  & -133 \\
 & rs2243552 & &  & 0.10 \scriptsize(0.03)& 0.03 \scriptsize(0.05)&  & -128 \\
 & rs2247809 & 0.01 \scriptsize(0.06)& 0.01 \scriptsize(0.06)& 0.18 \scriptsize(0.07)& 0.26 \scriptsize(0.06)& 0.01 \scriptsize(0.05)& -116 \\
 & rs878797$^{**}$ & 0.11 \scriptsize(0.06)& 0.11 \scriptsize(0.06)& 0.26 \scriptsize(0.07)& 0.23 \scriptsize(0.05)& 0.05 \scriptsize(0.06)& -55 \\
 & rs6516887 & 0.07  \scriptsize(0.04)& 0.07  \scriptsize(0.04)& -0.05  \scriptsize(0.07)& -0.09  \scriptsize(0.04)& 0.07  \scriptsize(0.04)& -44 \\
 & rs8128844 &  &  & -0.10 \scriptsize(0.06)& -0.17 \scriptsize(0.05)& 0.02 \scriptsize(0.04)& -24 \\
 & rs965951$^{**}$ & 0.10 \scriptsize(0.10)& 0.10 \scriptsize(0.10)& 0.28 \scriptsize(0.11)& 0.26 \scriptsize(0.08)& 0.13 \scriptsize(0.09)& -13 \\
 & rs2070610 & & & 0.18 \scriptsize(0.05)& 0.17 \scriptsize(0.04)&  & -0 \\
 & rs2832159 & 0.06 \scriptsize(0.06)& 0.06 \scriptsize(0.06)& -0.04 \scriptsize(0.07)& -0.20 \scriptsize(0.06)& 0.11 \scriptsize(0.05)& 13 \\
 & rs2832178$^{**}$ & -0.16  \scriptsize(0.06)& -0.16  \scriptsize(0.06)& -0.16  \scriptsize(0.08)& -0.20  \scriptsize(0.06)& -0.24  \scriptsize(0.06)& 34 \\
 & rs2832186 & & & -0.06  \scriptsize(0.07)& 0.12  \scriptsize(0.05)&  & 38 \\
 & rs2832190$^{**}$ & -0.41  \scriptsize(0.06)& -0.41  \scriptsize(0.06)& -0.25  \scriptsize(0.11)& -0.20  \scriptsize(0.08)& -0.49  \scriptsize(0.06)& 42 \\
 & rs2832190 & -0.22  \scriptsize(0.05)& -0.22  \scriptsize(0.05)& -0.16  \scriptsize(0.05)& -0.26  \scriptsize(0.06)& -0.28  \scriptsize(0.05)& 42 \\
 & rs7275293 &  &  & 0.13  \scriptsize(0.12)& 0.32 \scriptsize(0.09)&  & 54 \\
 & rs16983792 & -0.11 \scriptsize(0.04)& -0.11 \scriptsize(0.04)& -0.10 \scriptsize(0.05)& -0.18 \scriptsize(0.04)& -0.14 \scriptsize(0.04)& 82 \\
 & rs2251381$^{**}$ &  & & -0.11 \scriptsize(0.08)& -0.15 \scriptsize(0.05)&  & 85 \\
 & rs2251517$^{**}$ & -0.25 \scriptsize(0.05)& -0.25 \scriptsize(0.05)& -0.26 \scriptsize(0.07)& -0.27 \scriptsize(0.05)& -0.28 \scriptsize(0.05)& 86 \\
 & rs2251517 & -0.11 \scriptsize(0.04)& -0.11 \scriptsize(0.04)& -0.19 \scriptsize(0.05)& -0.23 \scriptsize(0.03)& -0.15 \scriptsize(0.04)& 86 \\
 & rs2832225 &  &  &  & -0.07  \scriptsize(0.03)&  & 87 \\
 & rs7283854 &  &  & 0.10 \scriptsize(0.04)& 0.13 \scriptsize(0.02)&  & 443 \\
   \hline
\end{tabular}
\end{table}
The eQTLs are almost all located within 500kb upstream TSS or 500kb downstream TES (Figure 2) and mostly from 100kb upstream TSS to 350 kb downstream TES.
\begin{figure}
\includegraphics[width=\textwidth]{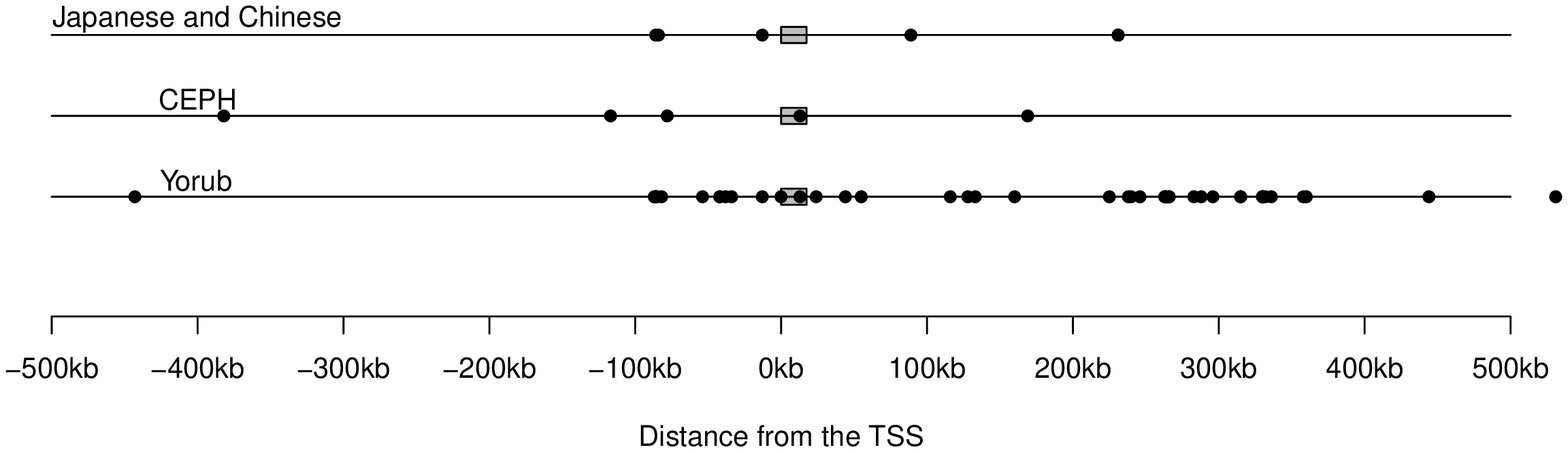}\label{fig:loci}
\caption{Chromosome locations of identified eQTLs of the gene CCT8
  with grey region as the CCT8's coding region. The eQTLs selected by any
  of the five methods are shown.}
\end{figure}

\section{Discussion}\label{sec6}

In this paper, a robust and efficient penalized quasi-likelihood
approach is introduced for model selection with NP-dimensionality.  It
is shown that such an adaptive learning technique has a strong oracle
property.  As  specific examples, two complementary methods of
penalized composite $L_1$-$L_2$ regression and weighted composite
quantile regression are introduced and they are shown to possess good
efficiency and model selection consistency in ultrahigh dimensional
space.  Numerical studies show that our method is adaptive to unknown
error distributions and outperforms LASSO \citep{T96} and equally
weighted composite quantile regression  \citep{ZY08}.

The penalized composite quasi-likelihood method can also be used in
sure independence screening \citep{FL08,FS10} or iterated version \citep{FSW09}, resulting in a robust variable screening and selection.  In this case, the marginal regression
coefficients or contributions will be ranked and thresholded \citep{FL08,FS10}.  It can also be applied to the aggregation problems of classification \citep{BRT09} where the usual $L_2$ risk function could be replaced with composite quasi-likelihood function.
The idea can also be used to choose the loss functions in machine learning. For example, one can adaptively combine the hinge-loss function in the support vector machine, the exponential loss in the AdaBoost, and the logistic loss function in logistic regression to yield a more efficient classifier.

\setcounter{equation}{0}  
   \renewcommand{\theequation}{A.\arabic{equation}}

\section*{Appendix A: Regularity Conditions} \label{secappendix}

Let $D_k$ be the set of discontinuity points of $\psi_k(t)$,  which is a subgradient of $\rho_k$. Assume that the distribution of error terms $F_\varepsilon$ is smooth enough so that
$F_\varepsilon(\cup_{k=1}^K D_k)=0$.  Additional regularity conditions on $\psi_k$ are needed, as in \cite{BRW92}.

\begin{condition}\label{cond1} The function $\psi_k$ satisfies
$E[\psi_k(\varepsilon_1+c)]=a_kc+o(|c|) \mbox{ as } |c| \to 0$, for
some $a_k >0$. For sufficiently small $|c|$,
$g_{kl}(c)=E[(\psi_k(\varepsilon_1+c)-\psi_k(\varepsilon_1))(\psi_l(\varepsilon_1+c)-\psi_l(\varepsilon_1))] $ exists and is continuous at $c=0$, where $k,l=1,...,K$.
The error distribution satisfies the following Cram\'{e}r condition:  \phantom{JOe}
$E \left| \psi_{\mathbf{w}}(\varepsilon_i) \right|^m\leq m! R K^{m-2} $, for some constants
$R$ and $K$.
\end{condition}

This condition implies that $E \psi_k(\varepsilon_i) = 0$, which is an
unbiased score function of parameter $\bbeta$. It also implies that
$E \partial \psi_k (\varepsilon_i) = a_k$ exists.  The following two
conditions are important for establishing sparsity properties of
parameter $\hat{\bbeta}_\bw$ by controlling the penalty weighting scheme $\bd$ and the regularization parameter $\lambda_n$.

\begin{condition}\label{cond2} Assume that
$D_n=\max \{d_j:j \in \mathcal{M}_*\} =o(n^{\alpha_1 - \alpha_0/2})$ and $\lambda_n D_n =O(n^{-(1+\alpha_0)/2})$.  In addition, $
\liminf \min\{d_j: j\in \mathcal{M}_*^c\}>0.$
\end{condition}

The first statement is to ensure that the bias term in Theorem~\ref{thm2} is negligible.  It is needed to control the bias due to the convex penalty. The second requirement is to make sure that the weights $\bd$ in the second part are uniformly large so that the vanishing coefficients are estimated as zero.  It can also be regarded as a normalization condition, since the actual weights in the penalty are $\{ \lambda_n d_j \}$.

The LASSO estimator will not satisfy the first requirement of
Condition \ref{cond2} unless $\lambda_n$ is small and $\alpha_1 >
\alpha_0/2$.  Nevertheless, under the sparse representation condition
\citep{ZY06}, \cite{FL09} show that with probability tending to one,
the LASSO estimator is model selection consistent with
$\|\hat{\bbeta}_1 - \bbeta_{1}^*\|_\infty = O( n^{-\gamma} \log n)$,
when the minimum signal $\beta_n^* = \min \{|\beta_j^{*}|, j \in
\mathcal{M}_*\}\geq n^{-\gamma} \log n$. They also show that the same
result holds for the SCAD-type  estimators under weaker conditions.
Using one of them as the initial estimator, the weight $d_j =
\gamma_\lambda(\hat{\beta}_j^{0}) / \lambda$ in (\ref{eq8}) would
satisfy Condition \ref{cond2}, on a set with probability tending to
one.  This is due to the fact that with $\gamma_\lambda(\cdot)$ given
by (\ref{eq6}), for $j \in \mathcal{M}_*^c$, $d_j =
\gamma_\lambda(0)/\lambda = 1$, whereas for $j \in \mathcal{M}_*$,
$d_j \leq \gamma_\lambda(\beta_n^*/2) / \lambda = 0$, as long as
$\beta_n^* \gg n^{-\gamma} \log n = O(\lambda_n)$.  In other words,
the results of Theorems~\ref{thm1} and \ref{thm2} are applicable to
the penalized estimator (\ref{eq8}) with data driven weights.

\begin{condition}\label{cond3}
 The regularization parameter $\lambda_n \gg n^{-1/2 + (\alpha_0 - 2\alpha_1)_+ /2 + \alpha_2 }$,  where parameter $\alpha_1$ is defined in Condition \ref{cond5} and $\alpha_2 \in [0, 1/2)$ is a constant, bounded by the restriction in Condition~\ref{cond4}.
\end{condition}

We use the following notation throughout the proof.
Let $\bB$ be a matrix.  Denote by $\lambda_{\min}(\bB)$ and $\lambda_{\max}(\bB)$ the minimum and maximum eigenvalue of the matrix $\bB$ when it is a square symmetric matrix.  Let $\|\bB\| = \lambda^{1/2}_{\max}(\bB^T \bB)$ be the operator norm and $\|\bB\|_\infty$  the largest absolute value of the elements in $\bB$.  As a result, $\|\cdot \|$ is the Euclidean norm when applied to a vector.  Define $ \|\bB\|_{2, \infty} = \max_{\|\bv\|_2 = 1} \| \bB \bv\|_\infty$.

\begin{condition}\label{cond4}
The matrix $\bS^T \bS$ satisfies $C_1 n \leq \lambda_{\min}(\bS^T \bS) \leq \lambda_{\max}(\bS^T \bS) \leq C_2 n$ for some positive constants $C_1,C_2$. There exists $ \xi > 0$ such that
\[
\sum_{i=1}^n (\|\bS_i \|/n^{1/2})^{(2+\xi)}\to 0,
\]
where $\bS_i^T$ is the $i$-th row of $\bS$.  Furthermore, assume that the design matrix satisfies $||\bX||_{\infty} = O(n^{1/2 - (\alpha_0 - 2\alpha_1)_+/2 - \alpha_2})$ and $\max_{j \not \in \mathcal{M}_*} \|\bX_j^*\|^2 = O(n)$, where $\bX_j^*$ is the $j$-th column of $\bX$.
\end{condition}

\begin{condition}\label{cond5}
Assume that
$$
\sup_{\boldsymbol\beta \in \mathcal{B}(\boldsymbol\beta^*_1,\beta_n^*) } \| \bQ \mbox{diag} \{ \partial \bpsi_\bw (\bbeta)\} \bS \|_{2, \infty} = O(n^{1-\alpha_1}).
$$
\[
\max_{\boldsymbol\beta \in \mathcal{B}(\boldsymbol\beta^*_1,\beta_n^*) } \lambda_{\min}^{-1} \left( \bS^T \mbox{diag} \{\partial \bpsi_\bw (\bbeta)\}
  \bS \right) = O_P(n^{-1}),
\]
where  $\mathcal{B}(\bbeta^*_1, \beta_n^*) $ is an $s$-dimensional ball
centered at $\bbeta^*_1$ with radius $\beta_n^*$ and $\mbox{diag}
(\partial \bpsi_\bw (\bbeta))$ is the diagonal matrix with $i$-th
element equal to $\partial \psi_\bw (Y_i - \bS_i^T \bbeta)$.
\end{condition}

\setcounter{equation}{0}  
   \renewcommand{\theequation}{B.\arabic{equation}}

\section*{Appendix B:  Lemmas} \label{secappendix} 

Recall that $\bX = (\bS, \bQ)$ and $\mathcal{M}_* = \{1, \cdots, s\}$ is the true model.

\begin{lemma}\label{lem6.1}  Under Conditions~\ref{cond2} and \ref{cond4},
the penalized quasi-likelihood $L_n(\bbeta)$ defined by \eqref{eq9} has a unique global minimizer $\hat\bbeta = (\hat{\bbeta}_1^T, \mathbf{0}^T)^T$, if
  \begin{equation}\label{eqB1}
 \sum_{i=1}^n  \psi_{\mathbf{w}} \left( Y_i - \mathbf{X}_i^T \hat{\bbeta} \right) \bS_i + n \lambda_n \bd_{\mathcal{M}_*} \circ \textrm{sgn}(\hat\bbeta_1) = \mathbf{0},
 \end{equation}
 \begin{equation}\label{eqB2}
 || \bz (\hat \bbeta)||_{\infty} < n \lambda_n,
 \end{equation}
where $
\bz(\hat \bbeta)= \bd_{\mathcal{M}_*^c}^{-1} \circ \sum_{i=1}^n   \psi_{\mathbf{w}} \left( Y_i - \mathbf{X}_i^T \hat{\bbeta}\right) \bQ_i
$, $\bd_{\mathcal{M}_*}$ and $\bd_{\mathcal{M}_*^c}$ stand for the subvectors of $\bd$, consisting of its first $s$ elements and the last $p-s$ elements respectively, and $\mbox{sgn}$ and $\circ$ (the Hadamard product) in \eqref{eqB1} are taken coordinatewise.
Conversely, if $\hat\bbeta$ is a global minimizer of  $L_n(\bbeta)$, then \eqref{eqB1} holds and \eqref{eqB2} holds with strict inequality replaced with non-strict one.
\end{lemma}

\noindent{\bf Proof of Lemma \ref{lem6.1}:}  Under conditions~\ref{cond2} and \ref{cond4}, $L_n(\bbeta)$ is strictly convex.
Necessary conditions \eqref{eqB1} and \eqref{eqB2} are  direct consequences of the Karush-Kuhn-Tucker conditions of optimality. The sufficient condition follows from similar arguments as those in the proof of Theorem 1 in \cite{FL09} and the strict convexity of the function $L(\bbeta)$.

\bigskip

\begin{lemma}\label{lem6.2} Under Conditions \ref{cond1}-\ref{cond5} we have that
$$
 \| \hat{\bbeta}^{\mathbf{o}} - \bbeta^*\|_2 =O_P(\sqrt{s/n} + \lambda_n \| \bd_0 \|),
$$
where $\bd_0$ is the subvector of $\bd$, consisting of its first $s$ elements.
\end{lemma}

\noindent{\bf Proof of Lemma \ref{lem6.2}:} Since $\hat{\bbeta}^{\mathbf{o}}_2 =  \bbeta^*_2=0$, we only need to consider the  sub-vector of the first $s$ components.
Let us first show the existence of the biased oracle estimator.  We can restrict our attention to the $s$-dimensional subspace $\{\bbeta \in \mathbb{R}^p : \bbeta_{\mathcal{M}_0^c}=\mathbf{0} \}$.  Our aim is to show that
\begin{equation}\label{eqn:b1}
P\left( \inf_{||\bu||= 1} L_n\left(\bbeta_1^* + \gamma_n\mathbf{u}, \bzero \right) > L_n(\bbeta^*)\right) \to 1,
\end{equation}
for sufficiently large $\gamma_n$.  Here, there is a minimizer inside the ball
$\| \bbeta_1 - \bbeta_1^* \| < \gamma_n$, with probability tending to one. Using the strict convexity of $L_n(\bbeta)$, this minimizer is the unique global minimizer.

By the Taylor expansion at $\gamma_n = 0$, we have
$$
    L_n\left(\bbeta^*_1 + \gamma_n\mathbf{u},\mathbf{0}\right) - L_n(\bbeta^*_1,\mathbf{0})= T_1 +T_2,$$
where
\begin{eqnarray} \nonumber
T_1 &=& - \gamma_n  \sum_{i=1}^n  \psi_\mathbf{w}(\varepsilon_i) \bS_i^T \bu +  \frac{1}{2} \gamma_n^2  \sum_{i=1}^n  \partial \psi_\mathbf{w}(\varepsilon_i - \bar{\gamma}_n \bS_i^T \bu) (\bS_i^T \bu)^2 \\ \nonumber &=& -I_1 + I_2\\ \nonumber
T_2 &=& n\lambda_n
\sum_{j=1}^s d_j (\bigl|\beta^{*}_j + \gamma_n u_j\bigl|-  \bigl|\beta^{*}_j\bigl|).
\end{eqnarray}
where $\bar \gamma_n \in [0, \gamma_n]$.  By the Cauchy-Schwarz inequality,
$$
|T_2| \leq n \gamma_n \lambda_n \|\bd_0\| \|\bu \| = n \gamma_n \lambda_n \|\bd_0\|.
$$
Note that for all $\|\bu \| = 1$, we have
$$
|I_1| \leq \gamma_n  \| \sum_{i=1}^n \psi_\bw(\varepsilon_i) \bS_i \|
$$
and
$$
 E \| \sum_{i=1}^n \psi_\bw(\varepsilon_i) \bS_i \| \leq  \left (
 E \psi_\bw^2(\varepsilon) \sum_{i=1}^n \|\bS_i\|^2 \right )^{1/2}
 = \left ( E \psi_\bw^2(\varepsilon) \mbox{tr}(\bS^T \bS)\right )^{1/2},
$$
which is of order $O(\sqrt{ns})$ by Condition~\ref{cond4}.  Hence, $I_1 = O_p(\gamma_n \sqrt{ns})$ uniformly in $\bu$.

Finally, we deal with $I_2$. Let $H_i(c) = \inf_{|v| \leq c } \{ \partial \psi_{\mathbf{w}} (\varepsilon_i -  v)\}$.  By Lemma 3.1 of \cite{P84}, we have
\begin{eqnarray} \nonumber
I_2 &\geq& \gamma_n^2 \sum_{i=1}^n H_i (\gamma_n |\bS_i^T\bu| ) (\bS_i^T \bu)^2  \\ \nonumber &\geq&  c \gamma_n^2 n,
\end{eqnarray}
for a positive constant $c$.  Combining all of the above results, we have with probability tending to one that
\begin{eqnarray*}
L_n\left(\bbeta^*_1 + \gamma_n\mathbf{u},\mathbf{0}\right) - L_n(\bbeta^*_1,\mathbf{0}) & \geq &  n \gamma_n \{c \gamma_n - O_P(\sqrt{s/n}) - \lambda_n \|\bd_0\|\},
\end{eqnarray*}
where the right hand side is larger than 0 when $\gamma_n = B (\sqrt{s/n} + \lambda_n \|\bd_0\|)$ for a sufficiently large $B >0$.  Since the objective function is strictly convex, there exists a unique minimizer $\hat{\bbeta}_1^o$ such that
$$
 \| \hat{\bbeta}_1^o - \bbeta_1^* \| = O_P (\sqrt{s/n} + \lambda_n \|\bd_0\|).
$$



\bigskip

\begin{lemma} \label{lem6.3}
Under the conditions of Theorem \ref{thm2},
\begin{equation}\label{eqB3}
  [\mathbf{b}^T \bA_n \mathbf{b}]^{-1/2}\sum_{i=1}^n  \psi_{\mathbf{w}}(\varepsilon_i)
   \bb^T\bS_i  \stackrel{\mathcal{D}}{\to} \mathcal{N}(0,1)
\end{equation}
where $\bA_n=   E \psi_{\mathbf{w}}^2(\varepsilon) \bS^T \bS$ .
\end{lemma}

\noindent{\bf Proof of Lemma \ref{lem6.3}:}  By Condition~\ref{cond1}, since $\bS_i$ is independent of $\psi_{\mathbf{w}} (\varepsilon_i)$,
we have $E   \psi_{\mathbf{w}}(\varepsilon_i) \bS_i=0$, and
\begin{eqnarray}
 \mbox{Var} \left[   [\mathbf{b}^T \bA_n \mathbf{b}]^{-1/2} \sum_{i=1}^n \psi_{\mathbf{w}}(\varepsilon_i) \mathbf{b}^T \bS_i   \right] = 1.
\end{eqnarray}
To complete proof of the lemma, we  only need to check the Lyapounov condition.  By Condition~\ref{cond1}, $E |\psi_{\mathbf{w}} (\varepsilon)|^{2+\xi} < \infty$.  Furthermore, Condition~\ref{cond4} implies
$$
   \bb^T \bA_n \bb = E \psi_\bw ^2 (\varepsilon) \bb^T \bS \bS^T \bb \geq c_1 n,
$$
for a positive constant $c_1$.  Using these together with the Cauchy-Schwartz inequality, we have
\begin{eqnarray*}
& & \sum_{i=1}^n E \left| [\mathbf{b}^T \bA_n \mathbf{b}]^{-1/2}   \psi_{\mathbf{w}}(\varepsilon_i) \mathbf{b}^T \bS_i \right|^{2 + \xi}\\
& = & O(1)  \sum_{i=1}^n \left| n^{-1/2} \bb^T \bS_i  \right|^{2 + \xi}. \\
& = & O(1)  \sum_{i=1}^n \left| n^{-1/2} \|\bS_i\|  \right|^{2 + \xi},
\end{eqnarray*}
which tends to zero by Condition \ref{cond4}.  This completes the proof.

\bigskip

The following Bernstein's inequality can be found in Lemma 2.2.11 of \\
\citet{VW96}.

\begin{lemma}\label{lem6.4}
Let $Y_1,\cdots,Y_n$ be independent random variables with zero mean such that $E|Y_i|^m\leq
m!M^{m-2}v_i/2$, for every $m\geq 2$ (and all $i$) and some constants $M$ and $v_i$. Then
$$P\left(|Y_1+\cdots+Y_n|>t\right)\leq 2\exp\{-\frac{t^2}{2(v+Mt)}\},$$
for $v\geq v_1+\cdots v_n$.
\end{lemma}

Then the following inequality \eqref{eqB6} is a consequence of previous Bernstein's inequality. Let $\{Y_i\}$ satisfy the condition of Lemma~\ref{lem6.4} with $v_i \equiv 1$.
For a given sequence $\{a_i\}$, $E |a_i Y_i|^m \leq m! |a_i M|^{m-2} a_i^2/2$.  A direct application of Lemma~\ref{lem6.4} yields
\begin{equation}
P\left(|a_1 Y_1+\cdots+ a_n Y_n|>t\right)\leq 2\exp\{-\frac{t^2}{2( \sum_{i=1}^n a_i^2 + M
  \max_{i} |a_i| t)}\}.  \label{eqB6}
\end{equation}

\setcounter{equation}{0}  
   \renewcommand{\theequation}{C.\arabic{equation}}

  \section*{Appendix C: Proofs of Theorems} \label{secappendix} 

\noindent {\bf Proof of Theorem \ref{thm1}:}  We only need  to show
that $\hat{\bbeta}^o$ is the unique minimizer of $L(\bbeta)$ in
$\mathbb{R}^p$ on a set $\Omega_n$ which has a probability tending to
one.  Since $\hat \bbeta_1^o$ already satisfies  \eqref{eqB1}, we only
need  to check \eqref{eqB2}.

We now define the set $\Omega_n$. Let
$$
\bxi=(\xi_1,\cdots,\xi_p)^T= \sum_{i=1}^n   \psi_{{\mathbf{w}}} \left( Y_i - \bX_i^T \bbeta^* \right) \bX_i
$$
and consider the event $\Omega_n=\left\{ \left| \left|
      \bxi_{\mathcal{M}_*^c} \right| \right|_{\infty} \leq u_n
  \sqrt{n}\right\}$ with $u_n$ being chosen later. Then, by Condition~\ref{cond1} and Bernstein's inequality, it follows directly from \eqref{eqB6} that
$$
P \left\{\left| \xi_j  \right| > t \right\} \leq 2 \exp \left\{
-\frac{t^2}{2\left( \|\bX_j^* \|^2 R + t K \|\bX_j^*||_{\infty}\right)}\right\},
$$
where $\bX^*_j$ is the $j$-th column of $\bX$. Taking $t = u_n \sqrt{n}$, we have
\begin{equation}\label{eqC1}
P \left\{\left| \xi_j  \right| > u_n \sqrt{n} \right\} \leq 2 \exp \left\{
-\frac{u_n^2}{2\left( R \|\bX_j^* \|^2/n  +  K u_n \|\bX_j^*||_{\infty}/\sqrt{n}\right)}\right\} \leq e^{-c u_n^2},
\end{equation}
for some positive constant $c>0$, by Condition~\ref{cond4}.
Thus, by using the union bound, we conclude that
$$
P(\Omega_n) \geq1-  \sum_{j\in \mathcal{M}_*^c} P \left\{ |\xi_j | > u_n \sqrt{n} \right\} \geq
 1-2 (p-s) e^{-c u_n^2}.
$$

We now check whether \eqref{eqB1} holds on the set $\Omega_n$.  Let $\bpsi_\bw(\bbeta)$ be the $n$-dimensional vector with the $i$-th element $\psi_\bw(Y_i - \bX_i^T \bbeta)$.  Then, by Condition~\ref{cond2}
\begin{eqnarray}\label{eqC2}
\|\mathbf{z}(\hat{\bbeta}^o)\|_{\infty} & \leq & \left \| \bd_{\mathcal{M}_*^c}^{-1} \circ \bxi_{\mathcal{M}_*^c} \right\|_{\infty} + \left\| \bd_{\mathcal{M}_*^c}^{-1} \circ \bQ^T [\bpsi_{\mathbf{w}}(\hat \bbeta^o) -  \bpsi_{\mathbf{w}}(\bbeta^*)]
 \right \|_\infty \nonumber \\
 & = & O\left (  n^{1/2} u_n
 +  \left\|  \bQ^T \mbox{diag} (\partial \bpsi_{\mathbf{w}}(\bv)) \bS (\hat \bbeta_{1}^o - \bbeta_1^*)
 \right\|_{\infty} \right )
\end{eqnarray}
where $\bv$  lies between $\hat{\bbeta}^o$ and $\bbeta_1^*$.  By Condition~\ref{cond5}, the second term in \eqref{eqC2} is bounded by
$$
  O(n^{1 - \alpha_1}) \| \hat \bbeta_{1}^o - \bbeta_1^* \| = O_P \{n^{1 - \alpha_1} (\sqrt{s/n} + \lambda_n \|\bd_0\|)\},
$$
where the  equality follows from Lemma~\ref{lem6.2}.  By the choice of parameters,
$$
(n\lambda_n)^{-1} \|\mathbf{z}(\hat{\bbeta}^o)\|_{\infty}
= O\{n^{-1/2} \lambda_n^{-1} (u_n + n^{(\alpha_0-2\alpha_1)/2}) + D_n n^{\alpha_0/2-\alpha_1}\} = o(1),
$$
by taking $u_n = n^{(\alpha_0 - 2 \alpha_1)_+/2 + \alpha_2}$.
Hence, by Lemma~\ref{lem6.1}, $\hat{\bbeta}^o$ is the unique global minimizer.

\bigskip

\noindent{\bf Proof of Theorem \ref{thm2}:} By Theorem~\ref{thm1},
$\hat{\bbeta}_{\bw1} = \hat{\bbeta}_1^o$ almost surely.  It follows
from Lemma~\ref{lem6.2} that
$$
   \| \hat{\bbeta}_{\bw1} - \bbeta_{1}^* \| = O_P \{ \sqrt{s} ( \lambda_n D_n  + 1/\sqrt{n}) \}.
$$
This establishes the first part of the Theorem.

Let $Q_n(\bbeta_1) = \sum_{i=1}^n  \psi_{\mathbf{w}} (Y_i-\bS_i^T \bbeta_{1} ) \bS_i$.  By Taylor's expansion at the point $\bbeta_1^*$, we have
$$
Q_n(\hat\bbeta_{\bw1}) = Q_n(\bbeta_{1}^*) + \partial Q_n(\bv) (\hat\bbeta_{\mathbf{w}1} - {\bbeta}_1^*),
$$
where $\bv$ lies between the points $\hat\bbeta_{\mathbf{w}1}$ and $\bbeta_1^*$ and
\begin{equation} \label{eqC3}
\partial Q_n(\bv) = - \sum_{i=1}^n  \partial \psi_{\mathbf{w}}  (Y_i-\bS_i^T \bv) \bS_i \bS_i^T.
\end{equation}
By Lemma~\ref{lem6.2}, $\|\bv - \bbeta_1^*\| \leq \| \hat \bbeta_{\bw1} - \bbeta_1^*\| = o_P(1)$.

By using \eqref{eqB2}, we have
$$
  Q_n (\hat \bbeta_{\mathbf{w}1}) + n  \lambda_n \bd_{0}\circ
\mbox{sgn}(\hat\bbeta_{\mathbf{w}1}) = 0,
$$
or equivalently,
\begin{equation} \label{eqC4}
   \hat\bbeta_{\mathbf{w}1} - \hat{\bbeta}_1^* = -\partial Q_n(\bv)^{-1} Q_n (\bbeta_{1}^*) - \partial Q_n(\bv)^{-1} n  \lambda_n \bd_{0}\circ
\mbox{sgn}(\hat\bbeta_{\mathbf{w}1}).
\end{equation}
Note that $\| \bd_{0}\circ \mbox{sgn}(\hat\bbeta_{\mathbf{w}1}) \| = \| \bd_0\|$. We have for any vector $\bu$,
$$
   \left | \bu^T \partial Q_n(\bv)^{-1} \bd_{0}\circ
\mbox{sgn}(\hat\bbeta_{\mathbf{w}1})\right | \leq \| \partial Q_n(\bv)^{-1} \| \cdot \|\bu\| \cdot \|\bd_0\|.
$$
Consequently, for any unit vector $\bb$,
\begin{eqnarray*}
\left \| \bb^T (\bS^T \bS)^{1/2} \partial Q_n(\bv)^{-1} \bd_{0}\circ
\mbox{sgn}(\hat\bbeta_{\mathbf{w}1})\right \| &\leq& \lambda_{\max}^{1/2} (\bS^T \bS)
  \lambda_{\min}^{-1} (\partial Q_n(\bv)) \sqrt{s} D_n\\
  & = & O_P( \sqrt{s/n} D_n),
\end{eqnarray*}
by using Conditions~\ref{cond4} and \ref{cond5}.  This shows that the second term in (\ref{eqC4}), when multiplied by the vector $\bb^T (\bS^T \bS)^{1/2}$ is of order
$$
    O_P( \sqrt{sn} \lambda_n D_n) = o_P(1),
$$
by Condition~\ref{cond2}.  Therefore, we need to establish the asymptotic normality of the first term in \eqref{eqC4}.  This term is identical to the situation dealt by  \cite{P85}.  Using his result, the second conclusion of Theorem~\ref{thm2} follows.   This completes the proof.

\bigskip

\noindent{\bf Proof of Theorem \ref{thm3}:} First of all, by Taylor expansion,
\begin{eqnarray} \label{eqC5}
  \Phi_{n, \bw} (\hat{\bbeta}_1) = \Phi_{n, \bw} (\bbeta_1^*) + \Omega_{n, \bw} (\bar \bbeta_1) (\hat{\bbeta}_1 - \bbeta_1^*),
\end{eqnarray}
where $\bar{\bbeta}_1$ lies between $\bbeta_1^*$ and $\hat{\bbeta}_1$.  Consequently,
$$
\|\bar{\bbeta}_1 - \hat{\bbeta}_1\| \leq  \|\bbeta_1^* - \hat{\bbeta}_1\| = o_P(1).
$$
By the definition of the one step estimator \eqref{eq14} and \eqref{eqC5}, we have
\begin{eqnarray}\label{eqC6}
 \hat{\bbeta}^{\footnotesize{\mbox{os}}}_{\mathbf{w}1} - {\bbeta}^{*}_1
 =  \Omega_{n, \bw} (\hat \bbeta_1)^{-1}  \Phi_{n, \bw} (\bbeta_1^*) + \bR_n,
\end{eqnarray}
where
$$
\bR_n = \Omega_{n, \bw} (\hat \bbeta_1)^{-1}  \left \{ \Omega_{n, \bw} (\hat \bbeta_1)
- \Omega_{n, \bw} (\bar \bbeta_1) \right \} (\hat \bbeta_1 - \bbeta_1^*).
$$

We first deal with the remainder term.
Note that
\begin{equation}\label{eqC7}
\|\bR_n\| \leq \left \| \{\Omega_{n, \bw} (\hat \bbeta_1)\}^{-1} \right \| \cdot \left \|
\Omega_{n, \bw} (\hat \bbeta_1) - \Omega_{n, \bw} (\bar \bbeta_1) \right \| \cdot \|\hat{\bbeta}_1 - \bbeta_1^* \|
\end{equation}
and
\begin{equation} \label{eqC8}
\Omega_{n, \bw} (\hat \bbeta_1) - \Omega_{n, \bw} (\bar \bbeta_1) = \sum_{i=1}^n f_i(\hat{\bbeta}_1, \bar{\bbeta}_1) \bS_i \bS_i^T,
\end{equation}
where $f_i(\hat{\bbeta}_1, \bar{\bbeta}_1) = \partial \psi (Y_i - \bS_i^T \hat{\bbeta}_1)
- \partial \psi (Y_i - \bS_i^T \bar{\bbeta}_1)$.  By the Liptchiz continuity, we have
$$
  |f_i(\hat{\bbeta}_1, \bar{\bbeta}_1)| \leq C \|\bS_i\| \cdot \| \hat{\bbeta}_1 - \bar{\bbeta}_1 \|,
$$
where $C$ is the Liptchiz coefficient of $\partial \psi_{\bw} (\cdot)$.  Let $\bI_s$ be the identity matrix of order $s$ and $b_n = \lambda_{\max} \{\sum_{i=1}^n \| \bS_i \| \bS_i \bS_i^T\}$. By (\ref{eqC8}), we have
$$
\Omega_{n, \bw} (\hat \bbeta_1) - \Omega_{n, \bw} (\bar \bbeta_1)
 \leq  C \|\hat{\bbeta}_1 - \bar \bbeta_1\| \sum_{i=1}^n \| \bS_i \| \bS_i \bS_i^T
 \leq C \|\hat{\bbeta}_1 - \bar \bbeta_1\| b_n \bI_s.
$$
Hence, all of the eigenvalues of the matrix is no larger than $C \|\hat{\bbeta}_1 - \bar \bbeta_1\| b_n$.  Similarly, by (\ref{eqC8}),
$$
\Omega_{n, \bw} (\hat \bbeta_1) - \Omega_{n, \bw} (\bar \bbeta_1)
 \geq  -C \|\hat{\bbeta}_1 - \bar \bbeta_1\| \sum_{i=1}^n \| \bS_i \| \bS_i \bS_i^T
 \geq - C \|\hat{\bbeta}_1 - \bar \bbeta_1\| b_n \bI_s,
$$
and all of its eigenvalue should be at least $-C \|\hat{\bbeta}_1 - \bar \bbeta_1\| b_n$.
Consequently,
$$
   \left \|
\Omega_{n, \bw} (\hat \bbeta_1) - \Omega_{n, \bw} (\bar \bbeta_1) \right \|
\leq C \|\hat{\bbeta}_1 - \bar \bbeta_1\| b_n.
$$

By Condition~\ref{cond5} and the assumption of $\hat{\bbeta}_1$, it follows from (\ref{eqC7}) that
$$
    \|\bR_n\| = O_P (s/n \cdot b_n / n) = O_P(s^{3/2}/n).
$$
Thus, for any unit vector $\bb$,
$$
\bb^T (\bS^T\bS)^{1/2} \bR_n \leq \lambda_{\max}^{1/2}(\bS^T\bS) \|\bR_n\| = O_P(s^{3/2}/n^{1/2}) = o_P(1).
$$
The main term in \eqref{eqC6} can be handled by using Lemma~\ref{lem6.3} and the same method as \cite{P85}. This completes the proof.

\begin{singlespace}

\end{singlespace}

\end{document}